\documentclass[]{aa} 
\usepackage{times,natbib,graphicx,amssymb}
\usepackage[english]{babel}
\def\cii{\ion{C}{II}}
\def\hii{\ion{H}{II}}
\def\oi{\ion{O}{I}}

\begin{document}
\title{A spatially resolved study of photoelectric heating and [\cii]
  cooling in the LMC}
\subtitle{Comparison with dust emission as seen by SAGE}

\author{
  D. Rubin\inst{1} \and
  S. Hony \inst{1} \and
  S.~C. Madden \inst{1} \and
  A.~G.~G.~M Tielens \inst{2} \and
  M. Meixner\inst{3} \and
  R. Indebetouw \inst{4} \and
  W. Reach\inst{5} \and
  A. Ginsburg \inst{6} \and
  S. Kim \inst{7} \and
  K. Mochizuki \inst{8} \and
  B. Babler \inst{9} \and
  M. Block \inst{10}\and
  S.~B Bracker \inst{9} \and
  C.~W. Engelbracht\inst{10}\and
  B.--Q. For \inst{10}\and
  K. Gordon\inst{10}\and
  J.~L. Hora\inst{11} \and
  C. Leitherer \inst{3} \and
  M. Meade\inst{9}\and
  K. Misselt\inst{10} \and
  M. Sewilo \inst{3} \and
  U. Vijh \inst{3} \and
  B. Whitney\inst{12} 
}

\authorrunning{Rubin et al}
\titlerunning{[\cii] and dust emission in the LMC}

\offprints{D. Rubin (Sacha.Hony@cea.fr)} 

\institute{
  Service d'Astrophysique, CEA/Saclay, l'Orme des Merisiers, F-91191
  Gif-sur-Yvette, France \and
  Kapteyn Institute, P.O. Box 800, NL-9700 AV Groningen, Netherlands
  \and
  Space Telescope Science Institute, 3700 San Martin Way, Baltimore,
  MD 21218, USA \and
  Department of Astronomy, University of Virginia, PO Box 3818,
  Charlottesville, VA 22903, USA \and
  Spitzer Science Center, California Institute of Technology, 220-6,
  Pasadena, CA, 91125, USA \and
  Center for Astrophysics and Space Astronomy, University of Colorado,
  Boulder, CO, USA \and
  Dept. of Astronomy \& Space Science, Sejong University, KwangJin-gu,
  KunJa-dong 98, Seoul, 143-747, Korea \and
  Institute of Space and Astronautical Science, Yoshinodai 3-1-1,
  Sagamihara, Kanagawa 229, Japan \and
  University of Wisconsin, Madison, WI 53706, USA \and
  Steward Observatory, University of Arizona, 933 North Cherry Ave.,
  Tucson, AZ 85719, Steward Observatory, USA \and
  Center for Astrophysics, 60 Garden St., MS 67 , Harvard University,
  Cambridge, MA 02138, USA \and
  Space Science Institute, 308 Morningside Ave., Madison, WI 53716, USA
}

\date{received \today; accepted date}

\abstract {Photoelectric heating is a dominant heating mechanism for
  many phases of the interstellar medium. We study this mechanism
  throughout the Large Magellanic Cloud.}
{We aim to quantify the importance of the [\cii] cooling line and the
  photoelectric heating process of various environments in the LMC
  and to investigate which parameters control the extent of
  photoelectric heating.}
{We use the BICE [\cii] map and the Spitzer/SAGE infrared maps. We
  examine the spatial variations in the efficiency of photoelectric
  heating: photoelectric heating rate over power absorbed by grains,
  i.e. the observed [\cii] line strength over the integrated infrared
  emission. We correlate the photoelectric heating efficiency and the
  emission from various dust constituents and study the variations as
  a function of H$\alpha$ emission, dust temperatures, and the total
  infrared luminosity. The observed variations are
  interpreted in a theoretical framework. From this we estimate
  radiation field, gas temperature, and electron density.}
{We find systematic variations in photoelectric efficiency. The
  highest efficiencies are found in the diffuse medium, while the
  lowest coincide with bright star-forming regions ($\sim$1.4 times
  lower). The [\cii] line emission constitutes 1.32\% of the far
  infrared luminosity across the whole of the LMC. We find
  correlations between the [\cii] emission and ratios of the mid
  infrared and far infrared bands, which comprise various dust
  constituents. The correlations are interpreted in light of the
  spatial variations of the dust abundance and by the local
  environmental conditions that affect the dust emission properties.
  As a function of the total infrared surface brightness,
  $S_\mathrm{TIR}$, the [\cii] surface brightness can be described as:
  $\mathrm{S_\mathrm{[\cii]}=1.25~S_\mathrm{TIR}^{0.69}~[10^{-3}~erg~s^{-1}cm^{-2}sr^{-1}]}$,
  for
  $\mathrm{S_\mathrm{TIR}~\gtrsim~3.2\cdot~10^{-4}erg~s^{-1}cm^{-2}sr^{-1}}$.
  We provide a simple model of the photoelectric efficiency as a
  function of the total infrared. We find a power-law relation between
  radiation field and electron density, consistent with other studies.
  The [\cii] emission is well-correlation with the 8~$\mu$m emission,
  suggesting that the polycyclic aromatic hydrocarbons play a dominant
  role in the photoelectric heating process.}
{}

\maketitle

\section{Introduction}
\label{sec:intro}
The structure and evolution of the interstellar medium (ISM) is
largely dependent upon the thermal processes taking place
\citep{goldsmith:1969, dejong:1977, mckee:1977, draine:1978,
  ferriere:1988}. This, in turn, shapes the evolution of galaxies as a
whole, as the constituents of the ISM are responsible for the
characteristics of incipient stellar generations. Therefore, an
understanding of the agents which dominate the heating and cooling of
interstellar gas is of fundamental importance.

A dominant heating source of the ISM of galaxies is the
photoelectric (PE) emission of interstellar dust grains. Absorption of
a far-ultraviolet (FUV) photon by a dust grain may result in the
ejection of an energetic electron which heats interstellar gas via
collisions. Photoelectric emission as a heating mechanism of the ISM
was first proposed by \citet{spitzer:1948} and later revisited by
\citet{watson:1972} and \citet{dejong:1977}. Since then, it has been
found that the process dominates the heating of a range of
interstellar media: neutral atomic gas clouds, the photo-dissociation
regions (PDRs), and the warm inter-cloud medium \citep[e.g.][]{1982A&A...106....1M,weingartner:2001}

The PE heating process has received much theoretical attention
\citep[e.g.][]{watson:1972, dejong:1977, draine:1978, tielens:1985a,
  bakes:1994, dwek:1996, weingartner:2001}. Due to grain charging, PE
heating efficiency is highly dependent on the physical conditions
which determine the ionisation and recombination rates. Specifically,
it depends on the FUV radiation field, gas temperature and electron
density. In turn, the extent of grain charging and therefore the
efficiency of PE heating is also highly dependent on the grain species
involved.

The $\mathrm{C^{+}}$ fine structure transition
($\mathrm{^{2}P_{3/2}-^{2}P_{1/2}}$) is the dominant coolant of the
diffuse ionised and diffuse atomic gas as well as in PDRs
\citep{dalgarno:1972, tielens:1985a, tielens:1985b, stacey:1991,
  madden:1993, petuchowski:1993, heiles:1994}. Several reasons account
for its dominance: carbon is the fourth most abundant element and it
has an ionisation potential of 11.3 eV, less than that of H. This
$\mathrm{C^{+}}$ transition is also easy to excite as it has a
relatively low excitation temperature ($\sim$92K). Thus, it is able to
cool warm neutral gas ($\mathrm{T\cong30-10^{4}K}$) whereas other
species can not \citep{tielens:1985a, tielens:1985b, wolfire:1990}.
The efficiency of the $\mathrm{C^{+}}$ as a coolant is dependent upon
environment. When temperatures or densities are high, other lines,
primarily [\oi]~63$\mu$m, participate in the cooling process
\citep{tielens:1985a, tielens:1985b, hollenbach:1991}. The critical
density of the $\mathrm{C^{+}}$ transition is relatively low
($\mathrm{\sim 3 \cdot 10^{3}~cm^{-3}}$). At densities above the
critical density or temperature above 92\,K the cooling by the [\cii]
line saturates and its importance as coolant diminishes.

An observational study of PE heating and gas cooling requires
[\cii]\footnote{Throughout this paper, we utilise the following
  notation: we refer to the fine structure transition with
  $\mathrm{C^{+}}$ and to the emission line that it produces with
  [\cii].} and infrared (IR) observations, covering
  wavelengths tracing the variety of dust components in the ISM, and
ideally of sufficient spatial resolution to separate environments,
because of the dependence of the process on environment and
composition. Because of its proximity, the Large Magellanic Cloud
(\object{LMC}) is an obvious candidate for a study of these processes.
We undertake such a study using the Spitzer legacy program: Surveying
the Agents of a Galaxy's Evolution \citep[SAGE][]{meixner:2006,
  bernard:2008}, and the Balloon-borne Infrared Carbon Explorer
mission \citep[BICE][]{mochizuki:1994}. SAGE fully mapped the
\object{LMC} at high spatial resolution from 3 to 160$\mu$m while the
latter offers a [\cii] map of the entire \object{LMC}. These datasets
offer advantages over previous work because of their enhanced spatial
resolution, wavelength coverage, and sensitivity. Prior studies using
NASA's Kuiper Airborne Observatory (KAO) and the Long Wavelength
Spectrometre (LWS) aboard the Infrared Space Observatory (ISO)
examined [\cii] emission integrated across whole galaxies
\citep[e.g.][]{stacey:1991, madden:1993, malhotra:2001}, or in
specific regions within galaxies \citep[e.g.][]{ stutzki:1988,
  meixner:1992, poglitsch:1995, israel:1996, madden:1997}. There have
not been many [\cii] studies which probe the range of different phases
of the ISM. Moreover, even though PE efficiency is believed to be a
strong function of grain size, there have not been many prior
observational studies of the effect of PE heating due to distinct
grain populations. Most studies have considered the total PE heating
across all grain populations because they did not have access to bands
which trace the emission from distinct grain populations.

The composition of the ISM of the \object{LMC} makes it an interesting
laboratory because it has a low metallicity; $\mathrm{Z
  \cong0.3-0.5Z_{\sun}}$ \citet{westerlund:1997} and $\mathrm{Z
  \cong0.25Z_{\sun}}$ \citet{dufour:1984}. The presence of metals, in
the form of dust, is integral to the PE effect. How does the low
metallicity lower dust abundance affect PE heating?

Moreover, it is known from observational studies that the \object{LMC}
and other low metallicity galaxies have a dearth of polycyclic
aromatic hydrocarbons (PAHs) compared to Galactic values
\citep{sakon:2006, vermeij:2002, madden:2006, wu:2006, galliano:2007,
  engelbracht:2008}. This condition raises another question: since it
is found theoretically that PAHs should play the greatest
  role in the PE heating process, how is the PE heating affected in a
galaxy with a prominent dearth of PAHs?

In contrast to the metallicity argument given above, it is found from
observations that the $\mathrm{C^{+}}$ coolant plays an even more
important role at low metallicity than it does at high metallicity ,
contributing up to $\sim$10 times more to the far-infrared (FIR)
emission \citep{ poglitsch:1995, israel:1996, madden:1997}. The larger
relative strength has been explained by the clumpy nature of the ISM
in low metallicity galaxies. Observations of [\cii]/CO show that for
low metallicity galaxies, this ratio can be 10-30 times higher than
for normal galaxies or active galaxies \citep[e.g.][]{madden:2000}.
The far UV radiation penetrates deeper into the molecular cloud at low
Z, for the same $A_{V}$, leaving a smaller CO core and larger
$\mathrm{C^{+}}$ emitting envelopes. The result is a preponderance of
CO cores clumps with larger exposed surface area. As the
$\mathrm{C^{+}}$ resides primarily in the envelopes of the clouds the
increase in surface area results in a higher ratio of
$\mathrm{[\cii]/CO}$ emission (e.g. \citet{roellig:2007}).

The aim of this paper is to explore the qualitative behaviour and
observationally quantify the extent of PE heating and [\cii] cooling
in relation to environment. The format of this paper is as follows:
Sec.~\ref{sec:observations} reviews the observational details of the
data used in this study. Section~\ref{sec:convolved_data} presents data
treatment process and the final images.

Section~\ref{sec:Cii_across_LMC} quantifies the importance of the [\cii]
coolant globally across the \object{LMC}. In
Sec.~\ref{sec:diffuse_and_non_diffuse} we examine PE heating as a
function of environment based on the H$\alpha$ surface brightness and
Sec.~\ref{sec:Cii_within_LMC} explores the variations in [\cii]
emission within the \object{LMC} using the H$\alpha$ criterion and
dust temperature. To explore the correlation between [\cii] emission
and the emission from the various grain components in the
\object{LMC}, Sec.~\ref{sec:dist_of_grain_emission} compares the
spatial distribution of the [\cii] emission and the emission at
various Spitzer bands. The key-concepts of PE heating are reviewed in
Sec.~\ref{sec:PE_heating_and_eff}, which are applied in
Sec.~\ref{sec:pe_eff_and_phys_cond} to describe the PE heating as a
function of radiation field within the \object{LMC}. Using the
observed relation we calculate electron densities, and find a
correlation between radiation field and electron density
(Sec.~\ref{sec:model_for_eff}). The paper concludes with
Sec.~\ref{sec:dust_and_PE_heating}, an analysis of the dependence of
PE heating on grain population; we study the qualitative behaviour of
the extent of PE heating on grain population, and we also quantify the
contributions to PE heating from the various grain populations.

\section{Observations}
\label{sec:observations}
\begin{table*}
  \caption{Summary of Data Used in this Study}
  \begin{tabular}{l ||cccccc}
    \hline
    \hline
    Line/Band&[\cii]&8 $\mu$m& 24 $\mu$m& 70 $\mu$m&160 $\mu$m& H$\alpha$\\
    \hline
    Mission & BICE & SAGE & SAGE & SAGE & SAGE & SHASSA\\
    1$\sigma$ limit [$\mathrm{MJysr^{-1}}$] & - &  0.2 &  0.2 & 1.0 & 2.0 & -\\
    1$\sigma$ limit [$\mathrm{10^{-5}erg~s^{-1}cm^{-2}sr^{-1}}$] $^{a}$ & 0.47 &  3.0 & 0.7 &  1.0 & 0.8 & 1.21 $\cdot$ $10^{-2}$\\
    1$\sigma$ limit [$\mathrm{MJysr^{-1}}$] $^{a}$ & - & $4 \cdot 10^{-4}$  &  $4 \cdot 10^{-4}$ & $4 \cdot 10^{-3}$ & $2 \cdot 10^{-2}$ & -\\ 
    1$\sigma$ limit [$\mathrm{10^{-5}erg~s^{-1}cm^{-2}sr^{-1}}$]$^{a}$  & 0.47 &  $6\cdot10^{-3}$ & $1\cdot10^{-3}$ &  $5 \cdot 10^{-3}$ & $1 \cdot 10^{-2}$ & $4.45 \cdot 10^{-5}$ \\
    Survey Area & 6$^\circ$ $\cdot$ 10$^\circ$ & 7.1$^\circ$$\cdot$ 7.1$^\circ$ & 7.8$^\circ$$\cdot$ 7.8$^\circ$ & 7.8$^\circ$$\cdot$ 7.8$^\circ$ & 7.8$^\circ$$\cdot$ 7.8$^\circ$ & $13^{\circ}\cdot13^{\circ}$\\
    Beamsize [$^{\prime\prime}$] & 894 & 2 & 6 & 18 & 40 & 240\\
    Linear size [pc] & $\sim$225 & $\sim$0.5 & $\sim$1.5 & $\sim$4.5 & $\sim$10 & $\sim$60 \\
    Reference & [1] & [2] & [2] & [2] & [2] & [3]\\
    \hline
    \hline
  \end{tabular}
  \newline
  $^{a}$ Sigma values after full data treatment, as outlined in
  Sec.~\ref{sec:convolved_data}. 
  References: [1] \citet{mochizuki:1994}; [2] \citet{meixner:2006};
  [3] \citet{gaustad:2001}.
  \label{tab:summary_of_data}  
\end{table*} 
\begin{figure*}[!tp]
  \includegraphics[clip,width=12cm,angle=90]{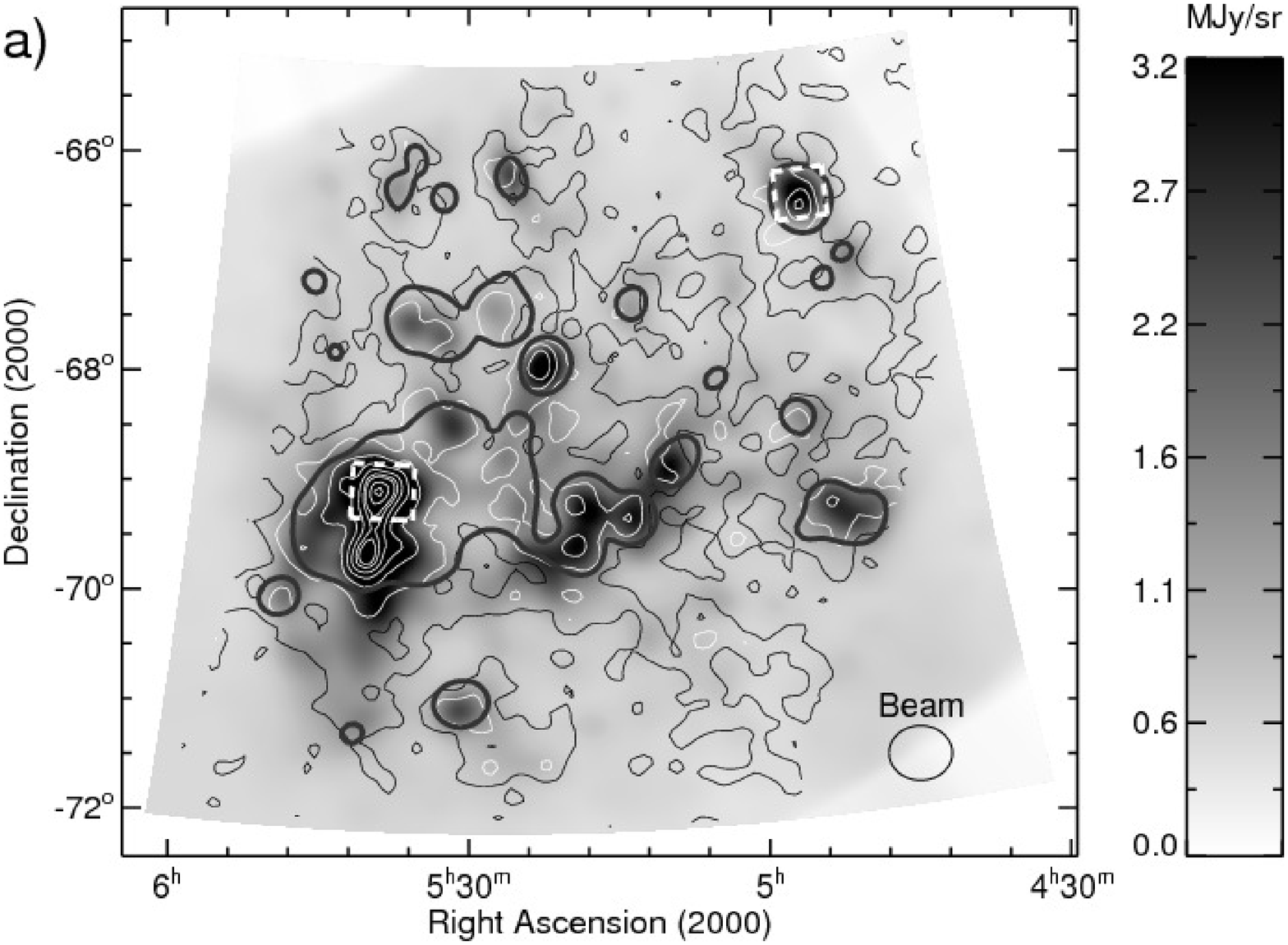} 
  \includegraphics[clip,width=12cm,angle=90]{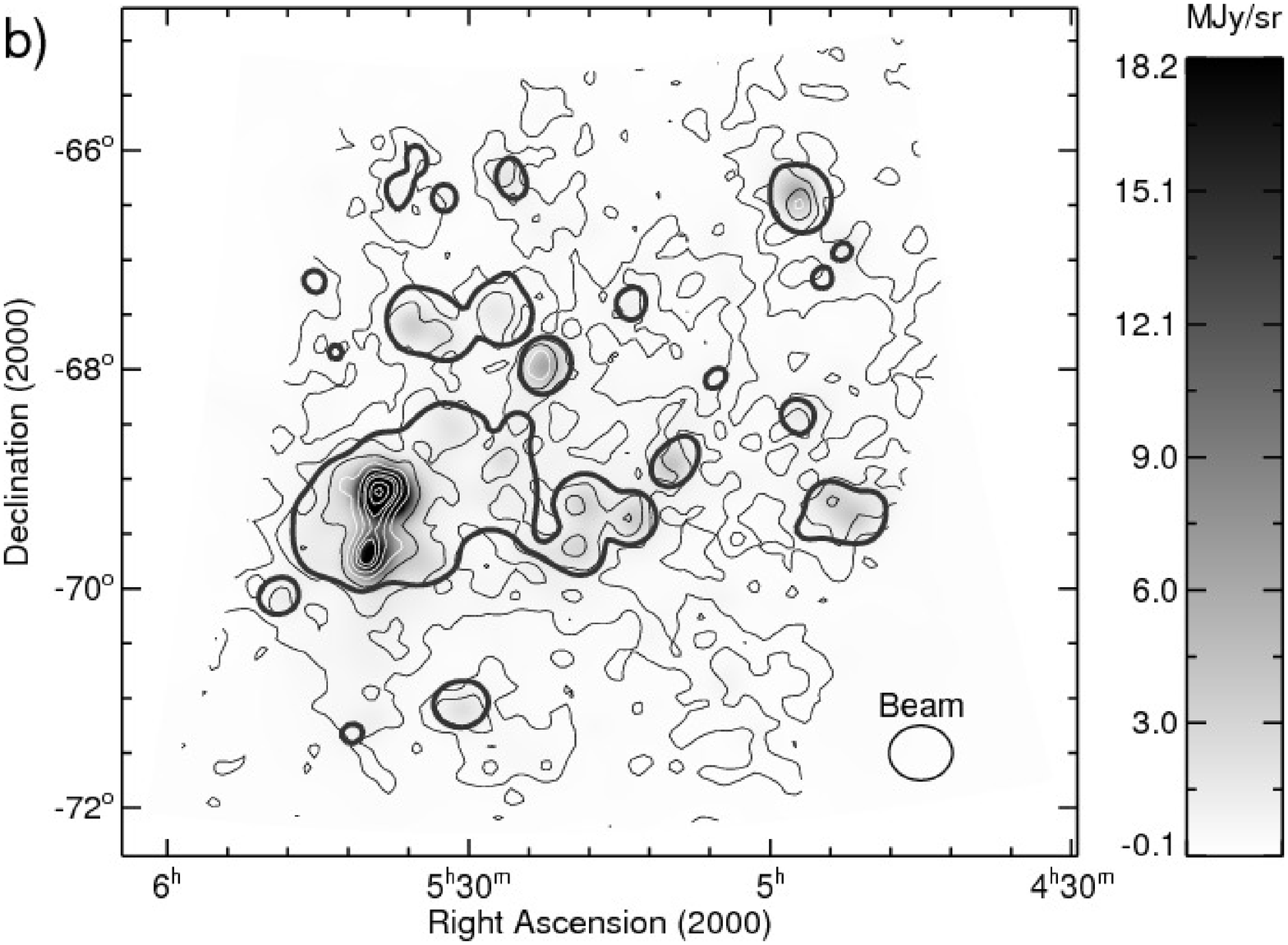}
  \caption{8$\mu$m (a) and 24$\mu$m (b) images convolved and
    re-gridded to the resolution and pixel scheme of the [\cii] map.
    To enhance the contrast of the low level emission, each image has
    been truncated at 0.25 times the max of that particular image.
    The [\cii] map is contoured over the images (thin grey and white
    contours) at levels of $\mathrm{1 \sigma_\mathrm{[\cii]}}$. The
    highest contours are white so that they can be distinguished from
    the underlying images. The thick grey lines enclose the regions
    which are defined as star forming given an H$\alpha$ surface
    brightness criterion (Sec.~\ref{sec:diffuse_and_non_diffuse}). The
    dashed white boxes in (a) indicate the regions used in the
    calculations for \object{30 Dor} (southeast) and \object{N11}
    (northwest) in Secs.~\ref{sec:Cii_within_LMC} and
    ~\ref{sec:dust_and_PE_heating}.}
  \label{fig:lmc_maps_1}
\end{figure*}
\begin{figure*}[!tp]
  \includegraphics[clip,width=12cm,angle=90]{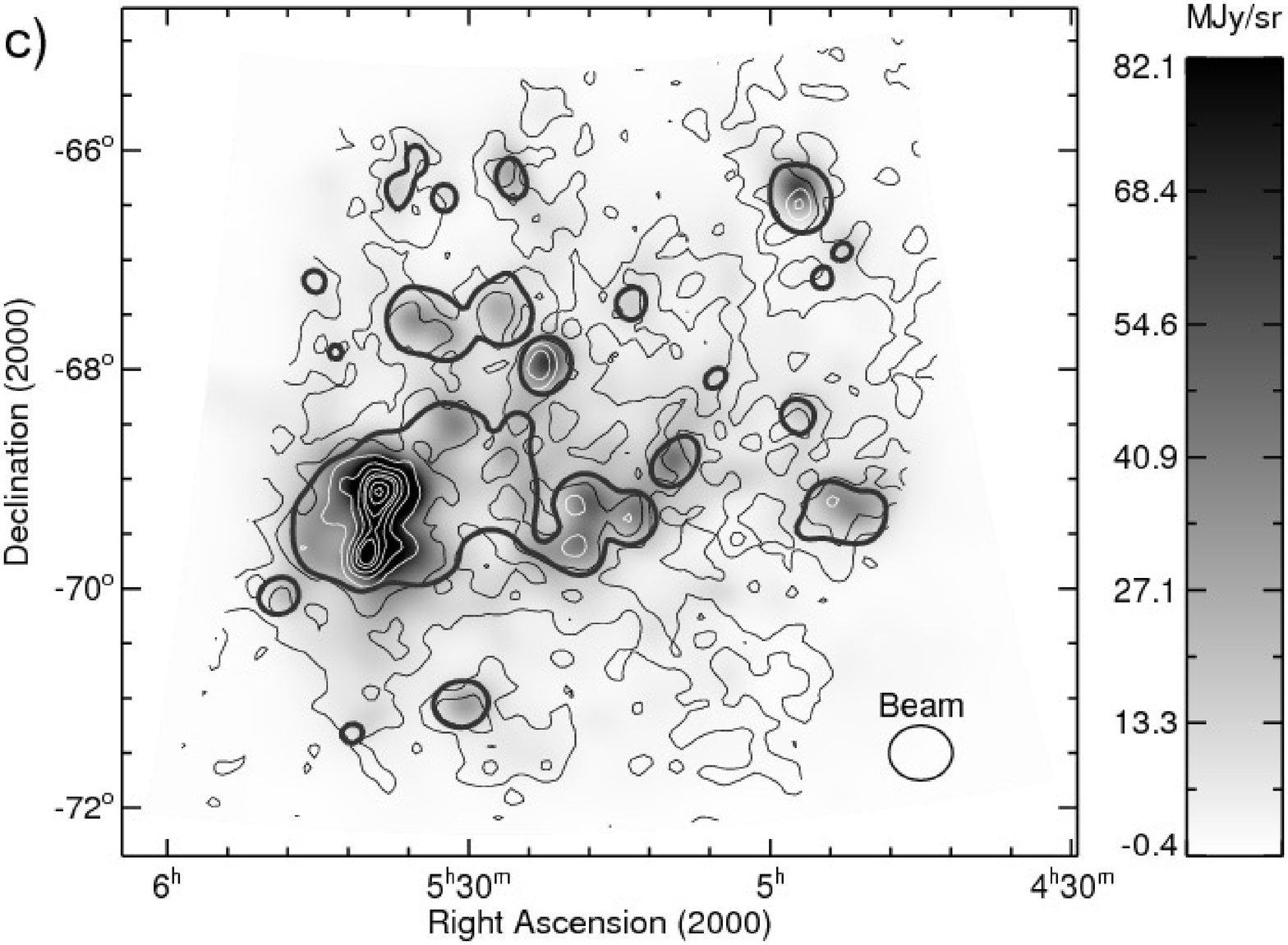} 
  \includegraphics[clip,width=12cm,angle=90]{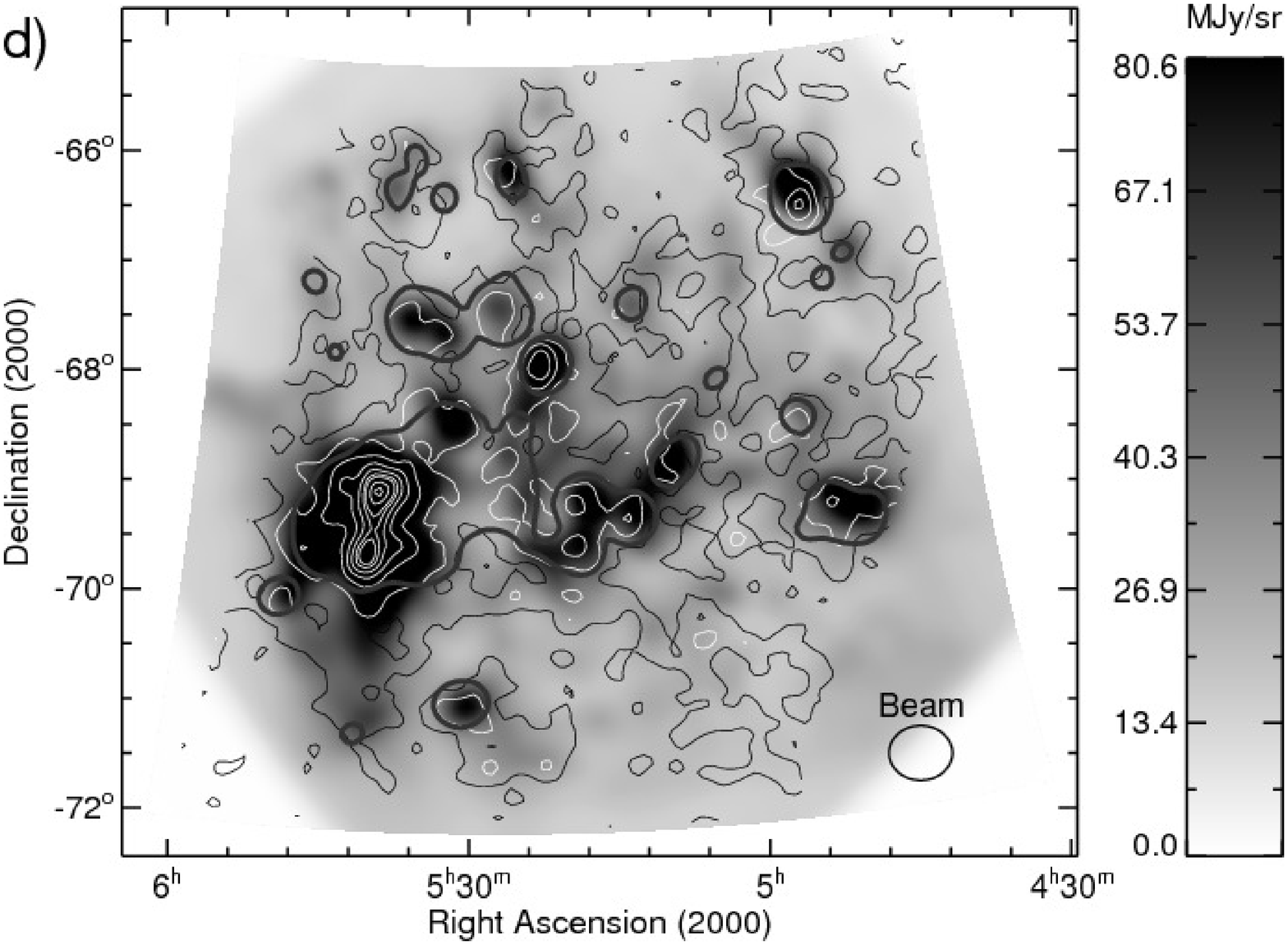}
  \caption{70$\mu$m (a) and 160$\mu$m (b) images convolved and
    re-gridded to the resolution and pixel scheme of the [\cii] map.
    To enhance the contrast of the low level emission, the images have
    been truncated at 0.25 times the max of that particular value.
    The [\cii] map is contoured over the images (thin grey and white
    contours) at levels of $\mathrm{1 \sigma_\mathrm{[\cii]}}$.
    The highest contours are white so that they can be distinguished
    from the underlying images. The thick grey lines enclose the
    regions which are defined as star forming given an H$\alpha$
    surface brightness criterion
    (Sec.~\ref{sec:diffuse_and_non_diffuse}).}
  \label{fig:lmc_maps_2} 
\end{figure*}
Relevant observational details of all the maps used in this study (the
[\cii], Spitzer and H$\alpha$ maps) are summarised in
Table~\ref{tab:summary_of_data}.

\subsection{The [\cii] line}
\label{sec:cii_data} The entire \object{LMC} was mapped in
  the 158$\mu$m [\cii] line by \citet{mochizuki:1994} during the BICE
mission. The velocity integrated [\cii] line surface
  brightness map was continuum subtracted by using a linear baseline
and foreground subtracted using COBE data to estimate the Milky Way
contribution. The measurements are calibrated against observations of
M17 carried out by \cite{matsuhara:1989}. We compare several regions
of the \object{30 Doradus} complex mapped by \cite{poglitsch:1995} and
\cite{israel:1996} aboard the KAO withe the BICE map and find
agreement better than $\sim20\%$. This is within the 30\% calibration
uncertainty of the BICE map as quoted by \citet{mochizuki:1994}. The
one $\sigma$ level for the BICE map as determined by
\cite{mochizuki:1994} is 0.47 $\cdot$ $10^{-5}\mathrm{erg~s^{-1}
  cm^{-2} sr^{-1}}$.

The BICE map offers a 6$^\circ$ $\times$ 10$^\circ$ field of view of
the \object{LMC} (see also Figs.~\ref{fig:lmc_maps_1}
and~\ref{fig:lmc_maps_2}). The beam has a FWHM of $14.9^{\prime}$,
corresponding to a linear size of $\sim225$pc at the distance of the
\object{LMC} \citep[50~kpc][]{feast:1999, keller:2006}. The positional
uncertainty is $\sim6^{\prime}$ \citep{mochizuki:1994}. We have
regrided the original map into pixels of $3^{\prime}$ ($\mathrm{\sim
  45pc}$) in length.

A significant fraction of the pixels are noise dominated: $46.1\%$,
$21.7\%$ and $10.8\%$ of all pixels in the final map are above are
above 1, 2 and 3 $\mathrm{3\sigma_\mathrm{[\cii]}}$, respectively. The
contours in Figs.~\ref{fig:lmc_maps_1} and~\ref{fig:lmc_maps_2} are
spaced at multiples of $\mathrm{\sigma_\mathrm{[\cii]}}$. To sample
the [\cii] emission from the diffuse ISM, in our analysis throughout
the rest of this paper, we consider pixels above the
$\mathrm{2\sigma_\mathrm{[\cii]}}$ level.

\subsection{Infrared Dust Emission}
\label{sec:sage_data}
To probe dust properties and abundances, we use the SAGE infrared and
mid-infrared survey of the \object{LMC} \citep{meixner:2006}. The
survey offers a $7^{\circ} \times 7^{\circ}$ field of view of the
\object{LMC} at effective wavelengths of 3.6, 4.5, 5.8, 8.0
\citep[IRAC][]{fazio:2004}, 24.0, 70.0 and 160.0$\mu$m
\citep[MIPS][]{rieke:2004}. We use the 8, 24, 70 and $160\mu$m bands
as the most important dust emission components can be traced primarily
by these bands.

Zodiacal and Milky Way contamination light at MIPS wavelengths has
been removed by doing off-source subtraction. The IRAC data have not
been background subtracted as the background contribution is
negligible compared to emission from the \object{LMC}. Averaging
several regions located off the \object{LMC} in the 8$\mu$m image
yields a background contribution of about $\mathrm{0.04 MJysr^{-1}}$.
Such a level is indeed small compared to the mean value of about
$\mathrm{0.46 MJysr^{-1}}$, for the 8$\mu$m image.

\subsection{H$\alpha$ Emission}
\label{sec:halpha_data}
As a general way to distinguish physical environments within the
\object{LMC} , we use an H$\alpha$ surface brightness criterion. The
H$\alpha$ map used is part of the Southern H-alpha Sky Survey Atlas
\citep{gaustad:2001}. The \object{LMC} is imaged in a single
$13^{\circ}\times13^{\circ}$ field and does not require mosaicing. The
images are continuum subtracted and smoothed to a resolution of
$4^{\prime}$. The sensitivity level of the H$\alpha$ maps is about 0.5
Rayleigh ( = 1.21 $\cdot$ $10^{-7}$ erg $\mathrm s^{-1}$ $\mathrm{
  cm^{2}}$ $\mathrm{sr^{-1}}$ at H$\alpha$).

\section{Data Treatment and Presentation}
\label{sec:convolved_data}
All IRAC, MIPS and H$\alpha$ data were convolved to the lowest
resolution data, the [\cii] map. The shape of [\cii] beam is not
precisely determined, and we therefore convolved the data with a
Gaussian kernel with a FWHM of $14.9^{\prime}$, the FWHM of the BICE
beam \citep{mochizuki:1994}. After convolution, the data were
interpolated to match the [\cii] pixel scheme.

To determine $\sigma$ values for the maps after the data treatment, we
carried out a facsimile of the data treatment process with simulated
Gaussian noise images. Standard deviations of the noise images after
the full data treatment were measured, and are included in
Table~\ref{tab:summary_of_data}. We use those $\sigma$ values in our
analysis.

Maps of the four Spitzer bands (convolved and re-gridded) with
overlays of the [\cii] contours are presented in
Figs.~\ref{fig:lmc_maps_1} and~\ref{fig:lmc_maps_2}. It is evident
from Figs.~\ref{fig:lmc_maps_1}a and~\ref{fig:lmc_maps_2}b that the
8$\mu$m and 160$\mu$m emission extends significantly into the bulk of
the galaxy. Figs.~\ref{fig:lmc_maps_1}b and~\ref{fig:lmc_maps_1}a show
that the 24$\mu$m and 70$\mu$m emission is much more concentrated
toward \hii~regions such as \object{30 Dor} and \object{N11}. More
attention to the spatial distributions of the grain emission will be
given in Sec.~\ref{sec:dist_of_grain_emission}.

Note the slight offsets of the [\cii] emission peaks from the IRAC and
MIPS emission peaks (especially near \object{N11}), which can be
larger than the 6$^{\prime}$ pointing accuracy. Similar deviations
were noted by \citet{kim:2002} in their analysis of atomic gas in
conjunction with the BICE map. We have carefully checked for problems
in the coordinate encoding used in the various maps. The observed
displacements do not appear to be due to such problems and thus we
conclude that there are real offsets between the peaks of emission of
[\cii] and IR emission, which can be as much as 100~pc. Indeed, the
region near \object{N11} is known to have most of its molecular
material in the direction of the displacement as compared to the
illuminating sources. Perhaps the observed displacement is due to
offsets in the peak emission at the molecular interface, as such
displacements are known to also occur in the Galaxy (e.g.
\citet{cesarsky:1996}).

\section{[\cii] Emission Globally Across the LMC}
\label{sec:Cii_across_LMC}
For comparison with other galaxies, we calculate the ratio of [\cii]
to the total infrared (TIR) integrated over the \object{LMC}. We adopt
the expression of \citet{dale:2002} to calculate the TIR from the
Spitzer the 24, 70 and 160$\mu$m filters, approximating the
  integrated 3 to 1000$\mu$m infrared surface brightness, $S_{TIR}$:
\begin{eqnarray}
  \label{eq:TIR_definition}
  \mathrm{S_\mathrm{TIR}}&=&\mathrm{1.559\nu S_{\nu,24}}+\mathrm{0.7686\nu S_{\nu,70}+1.347\nu S_{\nu,160}}.
\end{eqnarray}
The coefficients were derived from measured infrared SED shapes of a
sample of galaxies observed by Spitzer. \citet{draine:2007} also
provide an equation for the TIR luminosity using a modified
  prescription of the \citet{dale:2002} SED model and incorporating
  the 8$\mu$m IRAC band as well as the MIPS bands. The difference
between the two methods is less than $15\%$. Since both prescriptions
were made using SED properties from entire galaxies, it is important
to check that they yield good estimates on smaller scales as well. We
therefore compare the results from Eq.~(\ref{eq:TIR_definition}) with
a straight integration over all the IRAC and MIPS bands. We find no
systematic differences in the value of TIR as derived from the two
methods for distinct environments, i.e. SF and the diffuse medium.
Moreover, the values agree to within $\sim$10\%, which can most likely
be accounted for by the differences in the precise wavelength ranges
that both calculations consider.

The Spitzer definition of TIR emerged from the definition of
  $\mathrm{L_{FIR}}$ which was motivated by the IRAS bands
  \citep{helou:1986} and covers the FIR wavelength range from 42 to
  122$\mu$m. . The difference between TIR and FIR has been
observationally quantified by \citet{hunter:2001} who find that for
irregular galaxies (such as the \object{LMC}),
$\mathrm{L_\mathrm{TIR}/L_{FIR}} \approx2$, (with a dispersion of only
a few percent).

Integrated across the entire galaxy, we find that the total [\cii]
luminosity in the \object{LMC} is $\mathrm{(5.9 \pm
    1.8)\cdot 10^{6}L_{\sun}}$ assuming the distance to the LMC to
be 50\,kpc \citep{feast:1999, keller:2006}\footnote{The contribution
  of the [\cii] emission to the infrared emission is usually quoted in
  ratios of luminosities and we therefore adhere to this convention.
  We note, though, that we actually calculate values for this ratio
  with surface brightness. This is, however, equivalent to a ratio of
  luminosities assuming that both the [\cii] and TIR are radiated
  isotropically.}, consistent with \citet{mochizuki:1994} and
\citet{kim:2002}who estimate $\mathrm{L_\mathrm{[\cii]}}$ in the
\object{LMC} to be $\mathrm{5.7\cdot10^{6}L_{\sun}}$ and
$\mathrm{6.5 \cdot10^{6}L_{\sun}}$ respectively. We find that the
value of $\mathrm{L_\mathrm{[\cii]}/L_\mathrm{TIR}}$ is $0.46 \pm
0.14$\%. Assuming a factor of 2 between TIR and FIR,
$\mathrm{L_\mathrm{[\cii]}/L_{FIR}}=0.9\%$.

The relative contribution of [\cii] to the integrated FIR has often
been used to evaluate the global star formation activity in galaxies
\citep[e.g.][]{stacey:1991}. This value for the LMC is high compared
to normal and gas rich galaxies which normally have values of
$\mathrm{L_\mathrm{[\cii]}/L_{FIR}}$ less than 1\%. Values of
$\mathrm{L_\mathrm{[\cii]}/L_{FIR}}$ = .1\% to 0.5 \% are typical
\citep{stacey:1991, malhotra:1997, malhotra:2001}; for the Milky Way
$\mathrm{L_\mathrm{[\cii]}/L_{FIR}} \sim$0.3\% \citep{wright:1991}. Low metallicity galaxies can typically have
  $\mathrm{L_\mathrm{[\cii]}/L_{FIR}}$ as high as 1\% to 3\%
  \citep{poglitsch:1995, israel:1996, madden:1997, madden:2000}. This
  higher ratio is a consequence of the low metallicity: due to the
  reduced dust abundance, the overall mean free path of UV photons can
  be larger, resulting in a decrease in the FIR intensity arriving at
  the surfaces of the molecular clouds. To add to this effect, the
  lower dust attenuation results in the C$^{+}$-emitting regions being
  larger as the photo-dissociating photons traverse a larger volume of
  the molecular cloud.

\section{Distinction of Physical Environments}
\label{sec:diffuse_and_non_diffuse}
We aim to study the PE heating and the [\cii] cooling line as a
function of environment. While the spatial resolution of 225 pc
results in some mixing of phases, we can still delineate distinct
average conditions. We define two environments using the H$\alpha$
surface-brightness. H$\alpha$ emission is efficient at distinguishing
between dense \hii~regions and diffuse media. This is because
\hii~regions are H$\alpha$ bright due to their high free-electron
densities, which results in a high recombination rate. The H$\alpha$
line is a tracer of the physical parameters which determine the degree
of ionisation. That is, radiation field, temperature and density; all
of which play critical roles in PE heating and the $\mathrm{C^{+}}$
transition.

\citet{kim:2002} studied the \object{LMC} using the BICE [\cii] map
and made a similar distinction to separate the phases of the ISM. They
studied the $\mathrm{C^{+}}$ cooling rate for regions with an
H$\alpha$ surface brightness, $\mathrm{S_{H\alpha}}$, above and below
$\mathrm{4.25\cdot 10^{-5}erg~s^{-1} cm^{-2} sr^{-1}}$. This number
was based on the work of \citet{kennicutt:1986}. We considered using
this criterion, but this causes several well known \hii~regions such
as \object{N41}, \object{N144} and \object{N132} to be classified as
diffuse. We conclude that the criterion should be lowered to better
represent the different phases of the ISM. For a reformulation of the
criterion, we examined the data of \citet{kennicutt:1986}. They
photometrically observed \hii~regions in the \object{LMC} and
tabulated the H$\alpha$ surface brightness. A histogram of the
distribution of these surface brightnesses, shows the rapid fall off
at the lowest values, suggesting that the faint side of the
distribution is noise-dominated, while the bright side is comprised of
reliable values. Indeed, \citet{kennicutt:1986} warn that their
measurements of \hii~ regions with the lowest surface brightnesses are
unreliable. We therefore take the value of the peak of the
distribution, $\mathrm{1\cdot 10^{-5} erg~s^{-1} cm^{-2} sr^{-1}}$
, as the lowest reliable surface brightness for an \hii~region, which
we use to distinguish between physical environments in the
\object{LMC}. Every pixel with an H$\alpha$ surface brightness below
this value we call ``diffuse'', and every pixel with an H$\alpha$
surface brightness above it, we call ``star forming'' (SF).

The thick grey lines in Figs.~\ref{fig:lmc_maps_1}
and~\ref{fig:lmc_maps_2} enclose the SF pixels, while the diffuse
pixels reside outside the grey lines. One can see from this figure,
that the SF regions correspond to the brightest \hii~regions, such as
\object{30 Dor}, \object{N11} and the prominent \hii~regions along the
\object{LMC} bar, while not extending too far into the diffuse medium. While we can not avoid including diffuse emission within
  these SF regions, most SF pixels are dominated by \hii~regions and
  PDRs. Likewise the regions we label diffuse will undoubtedly contain
  some denser ionised material, but will be dominated by the diffuse
  conditions.

To validate the threshold, we estimate the electron density
($\mathrm{n_{e}}$) for the sets of SF and diffuse pixels under the
Case B approximation. The approximation provides
$\mathrm{S_{H\alpha}}$ which depends upon electron density, size of
the emitting regions and gas temperature \citep[e.g.][]{valls:1998}.
For the estimation, we use the mean values of $\mathrm{S_{H\alpha}}$
for the SF and diffuse pixels, and a temperature of $10^{4}$K. The
sizes of the emitting regions were determined by examining the
original H$\alpha$ images and visually determining the physical sizes
of typical SF~regions and the voids between them. For the SF pixels,
we assume that the light is dominated by the emission from
\hii~regions. The lengths determined were about 5$^{\prime}$ and
15$^{\prime}$ for the SF and diffuse regions respectively. For the SF
pixels, we assume that the light is dominated by the emission from
\hii~regions. This yields an average $\mathrm{n_{e}}$ of $\sim 100$
and $\mathrm{\sim 1cm^{-3}}$ for the SF and diffuse regions,
respectively. The latter density is reasonable for densities of the
warm ionised medium \citep{nordgren:1992}. The former is consistent
with \citet{peck:1997}, who find a mean electron density of $\sim 200$
cm$^{-3}$ in \object{30~Dor}.

\section{Variation of [\cii] Emission Within the LMC}
\label{sec:Cii_within_LMC}
\begin{figure}[!tp]
  \includegraphics[clip,height=8.8cm,angle=90]{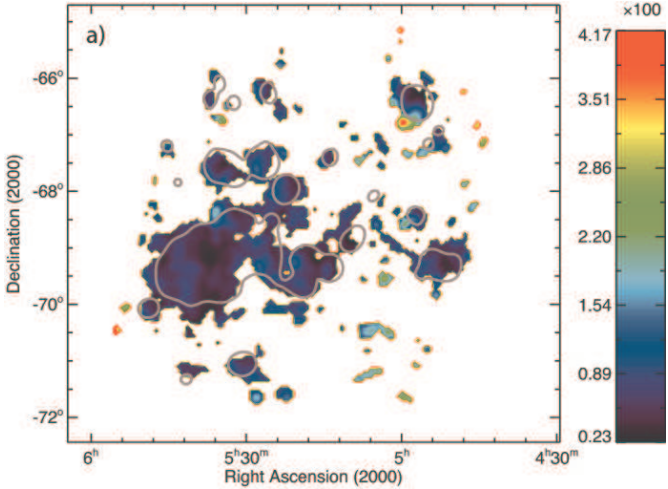} 
  \caption{Ratio map of
    L$_{\mathrm\mathrm{[\cii]}}$/L$_{\mathrm\mathrm{TIR}}$ (left).
    Pixels below 2$\sigma_{\mathrm\mathrm{[\cii]}}$ are set to white.
    The thick grey lines enclose the SF regions. (See online version
    for colour.)}
  \label{fig:map_efficiency}
\end{figure}
\begin{figure}[!tp]
  \includegraphics[clip,height=8.8cm,angle=-90]{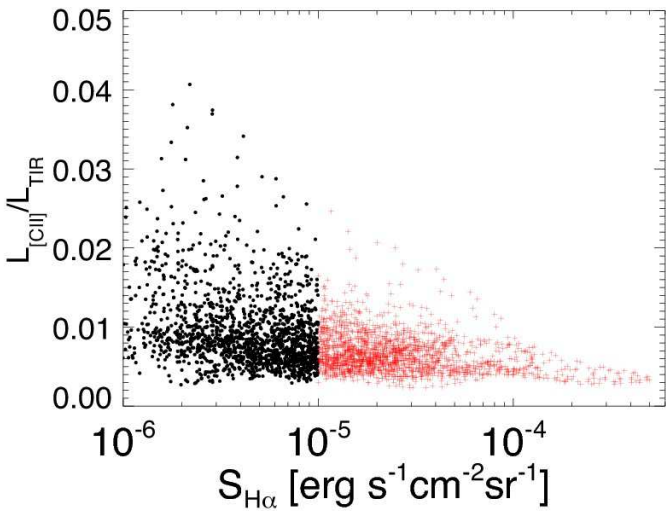}
  \caption{$\mathrm{L_{[\cii}/L_\mathrm{TIR}}$ for each pixel plotted
    against its H$\alpha$ surface brightness. The black dots and red
    crosses denote the diffuse and SF pixels respectively. (See online
    version for colour.)}
  \label{fig:eff_vs_T_and_Halpha} 
\end{figure}
%
\begin{table*}[!tp]
  \caption{Contribution of [\cii] from various regions in the LMC.}
  \begin{tabular}{lrrrrrr}
    \hline
    \hline
    Region                      &$\mathrm{L_\mathrm{[\cii]}/L_{[\cii],total}}$  &$\mathrm{Area/Area_{total}}$ &$\mathrm{L_\mathrm{TIR}/L_{TIR,total}}$  &$\mathrm{L_\mathrm{[\cii]}/L_{160}}$  &$\mathrm{L_\mathrm{[\cii]}/L_{[TIR]}^{a}}$ &$\mathrm{L_\mathrm{[\cii]}/L_{[FIR]}^{b}}$\\\\
    \hline
    Diffuse regions             &$51.8\pm0.1\%$                            &86\%                        & 59.0$\pm$ 0.1\% 	&0.032$\pm$0.1\%		                  &0.40$\pm$0.01\%		&0.80$\pm$0.01\%   \\
    SF regions                  &$48.2\pm0.1\%$                            &14\%                        &41.0 $\pm$ 0.1\%\		&0.063$\pm$0.1\%		          &0.55$\pm$0.01\%		&1.10$\pm$0.01\%   \\
    \object{30 Dor}             &$8.1\pm0.1\%$                             &0.8\%                       &9.0 $\pm$ 0.1\%		&0.072$\pm$0.1\%		          &0.42$\pm$0.01\% 		&0.84$\pm$0.01\%   \\
    \object{N11}                &$2.2\pm0.1\%$                             &0.7\%                       &2.0 $\pm$ 0.1\%		&0.058$\pm$0.1\%		          &0.57$\pm$0.01\%  		&1.14$\pm$0.03\%   \\
    total LMC			&$100\%$				   &$100\%$		        &$100\%$			&0.042$\pm$0.1\%		          &0.46$\pm$0.01\%   	        &0.92$\pm$0.02\%   \\
    \hline
    \hline
  \end{tabular}
  \label{tab:frac_of_CII}
  \newline
  $^{a}$ Also referred to as the photoelectric efficiency (see Sec.~7.2 ).
  $^{b}$FIR is defined using the IRAS bands (see section\ref{sec:sage_data}).
  TIR assumes the factor of 2 increase from the FIR, as found by \citet{hunter:2001} and discussed in the text (section \ref{sec:sage_data}). Due to the calibration uncertainties of the BICE map, all of these
  values have a {\it systematic} uncertainty of $\pm$30\%.
  The uncertainties in the table reflect the random noise which is
  important for comparing the relative.
 
\end{table*} 
To examine the variations of
$\mathrm{L_\mathrm{[\cii]}/L_\mathrm{TIR}}$, we show a ratio map of
$\mathrm{L_\mathrm{[\cii]}/L_\mathrm{TIR}}$ overlaid with the
boundaries of the SF regions in Fig.~\ref{fig:map_efficiency}. The
pixels with the lowest $\mathrm{L_\mathrm{[\cii]}/L_\mathrm{TIR}}$ are
associated with the centres of the brightest SF regions, confirming
the scenario in which the PE heating is least efficient at high
density.

The distribution of [\cii] and TIR from the different
phases are summarised in Table~\ref{tab:frac_of_CII}. Although the SF
regions have much higher values of [\cii] surface brightnesses,
approximately half of the [\cii] emission originates from the diffuse
medium. The SF regions contribute just less half of the [\cii\
luminosity ($\sim$48\%).

Fig.~\ref{fig:eff_vs_T_and_Halpha} shows
$\mathrm{L_\mathrm{[\cii]}/L_\mathrm{TIR}}$ as a function of the
H$\alpha$ surface brightness. As can be seen, the efficiency is roughly
constant across the range of H$\alpha$ surface-brightnesses. There are
some deviant pixels with high
$\mathrm{L_\mathrm{[\cii]}/L_\mathrm{TIR}}$, in particular at the
lowest values of $S_{\mathrm{H}\alpha}$, which are most-likely the
result of the noise in the [\cii] map. The bulk of our points clutter
around $\mathrm{L_\mathrm{[\cii]}/L_\mathrm{TIR}}$ $\simeq$ 0.005,
which is similar to the values found by \citet{malhotra:2001} and
\citet{hunter:2001}, who examine spiral and irregular galaxies
(including the \object{LMC}). There is evidence that on the whole the
\object{30 Dor} region has a slightly lower ratio of
$\mathrm{L_\mathrm{[\cii]}/L_\mathrm{TIR}}$ than the rest of the LMC
by less than 10\% (see also Fig.~\ref{fig:map_efficiency}).

Theoretically, such a decrease in this ratio towards the densest
regions is expected. The PDR models of \citet{tielens:1985a,
  tielens:1985b, wolfire:1990} show that [\cii] emission levels off at
the highest gas temperatures, radiation fields and gas densities.
Other lines take over (part of) the cooling process at high gas
temperature and high gas density. Since dust temperature roughly
scales with gas temperature and gas density, it is possible that the
observed decrease is associated with the critical temperature and
critical density of the $\mathrm{C^{+}}$ transition being reached. We,
however, do examine alternative explanations in the following
sections.

\section{Distribution of [\cii] Emission and Grain Component Emission in the LMC}
\label{sec:dist_of_grain_emission}
\begin{figure*}[!tp]
  \includegraphics[clip,width=8.8cm,angle=0]{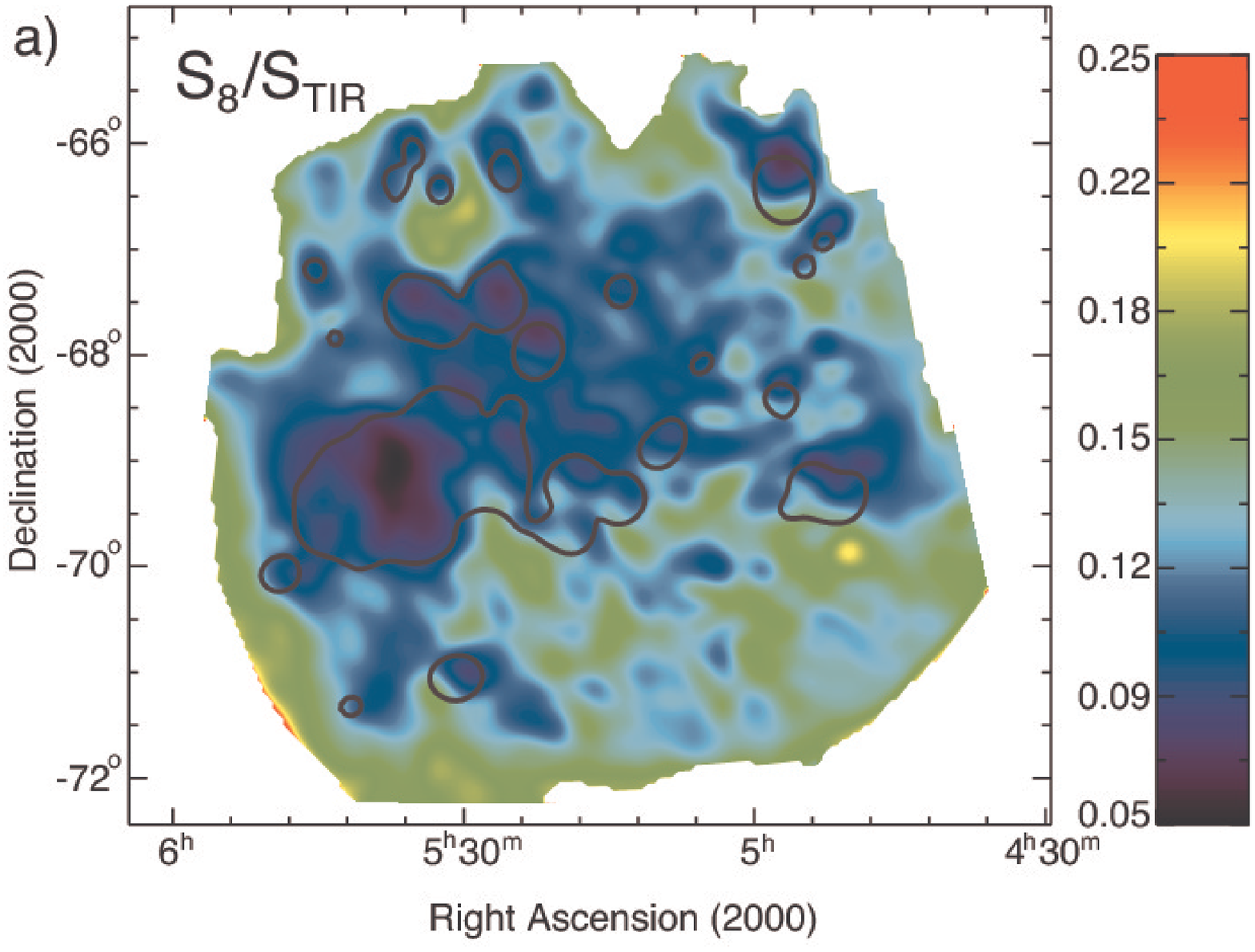}
  \includegraphics[clip,width=8.8cm,angle=0]{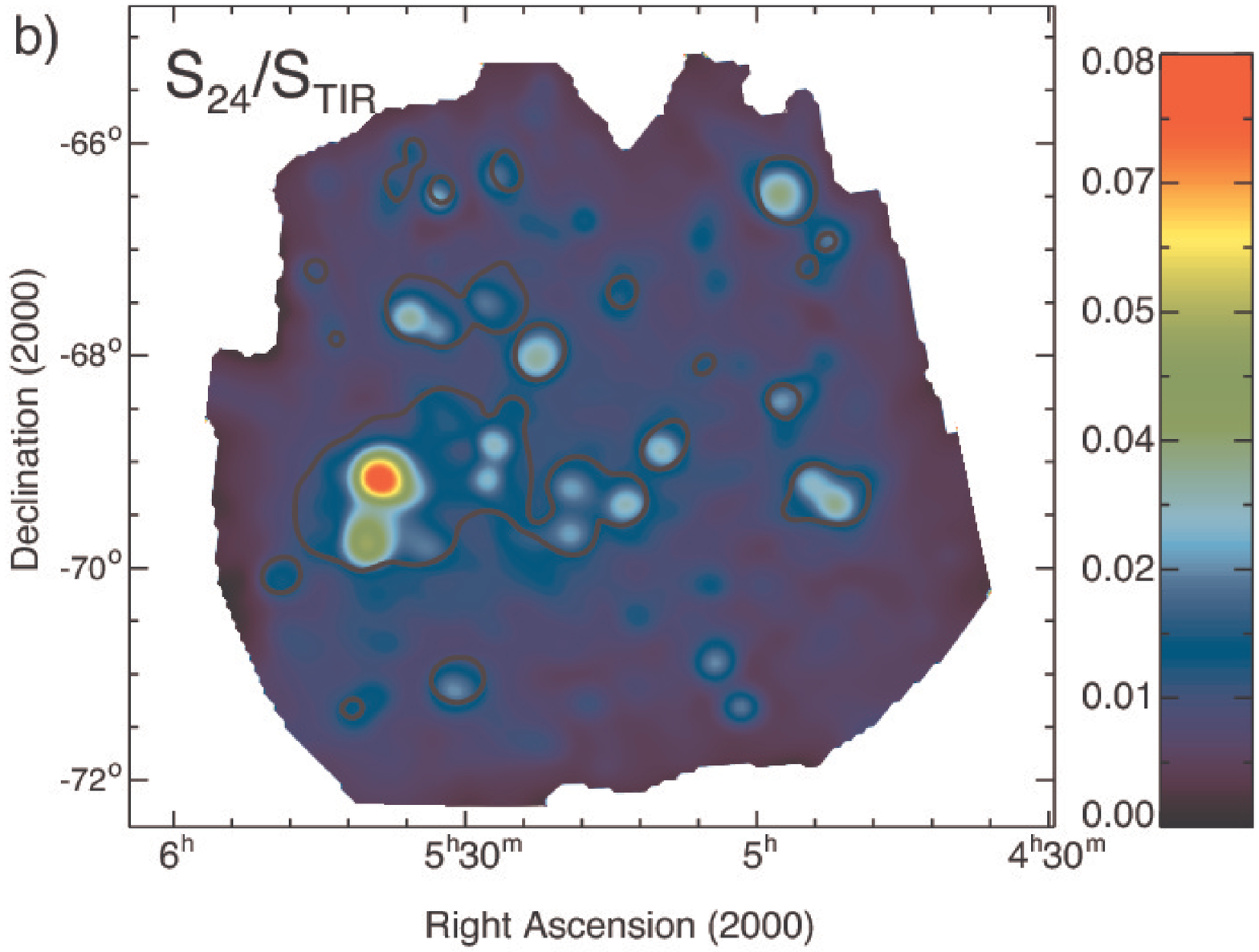}
  \includegraphics[clip,width=8.8cm,angle=0]{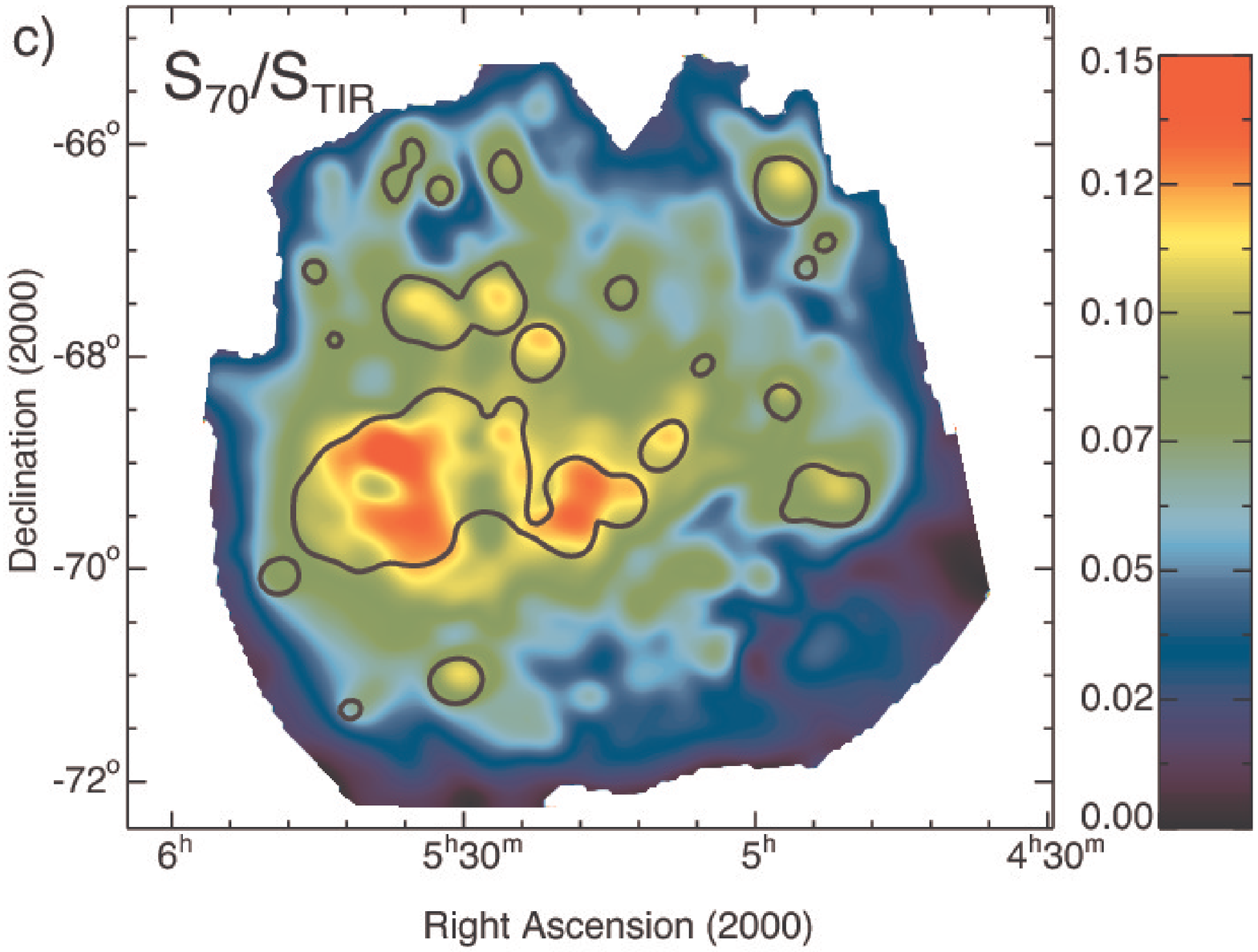}
  \includegraphics[clip,width=8.8cm,angle=0]{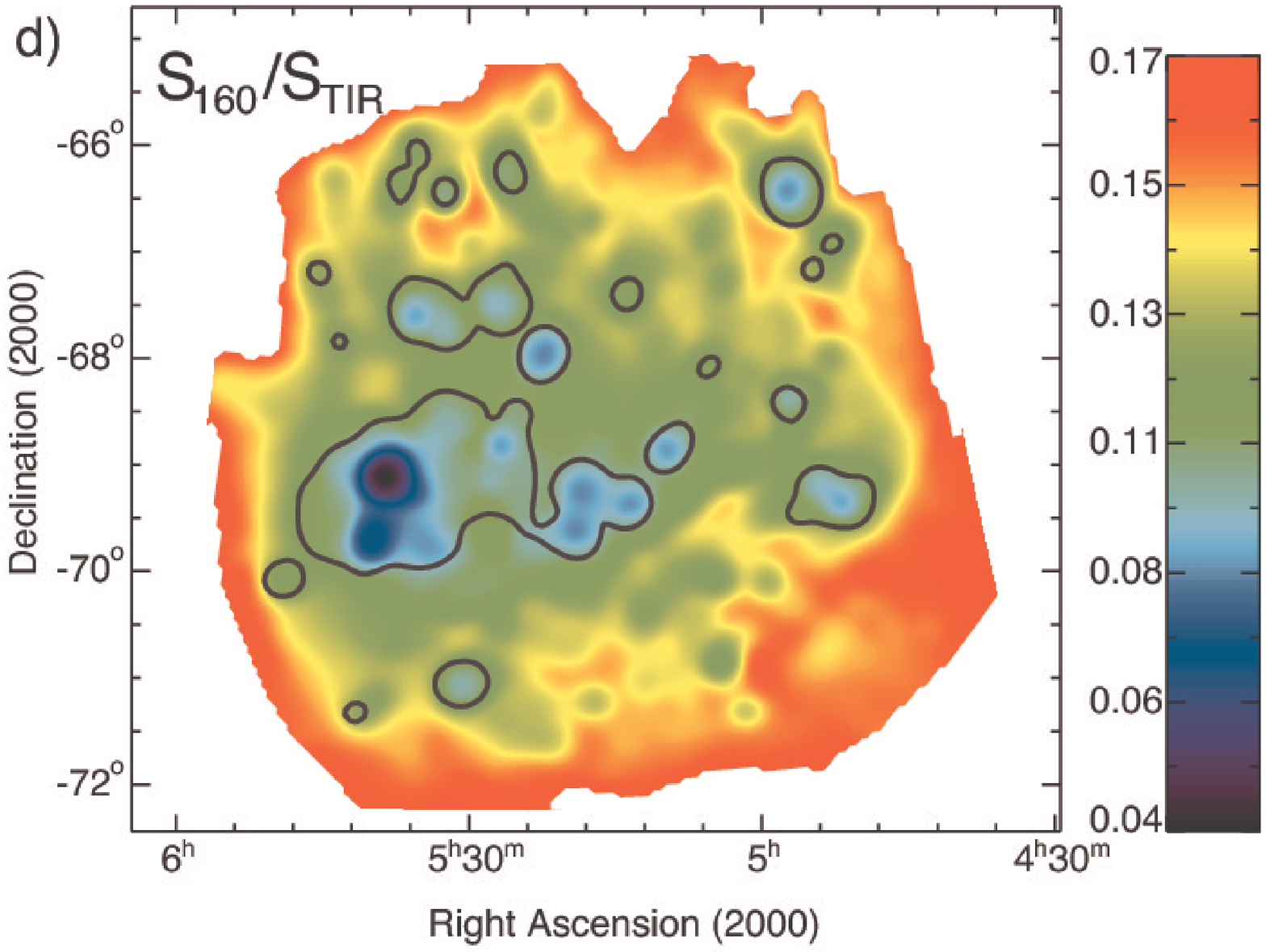}
  \caption{Ratio maps of the 8 (a), 24 (b), 70 (c) and 160$\mu$m (d)
    bands to the TIR emission. Contours enclose the SF regions. (See
    online version for colour.)}
  \label{fig:ratio_maps} 
\end{figure*}
\begin{figure*}[!tp]
  \includegraphics[clip,width=\textwidth]{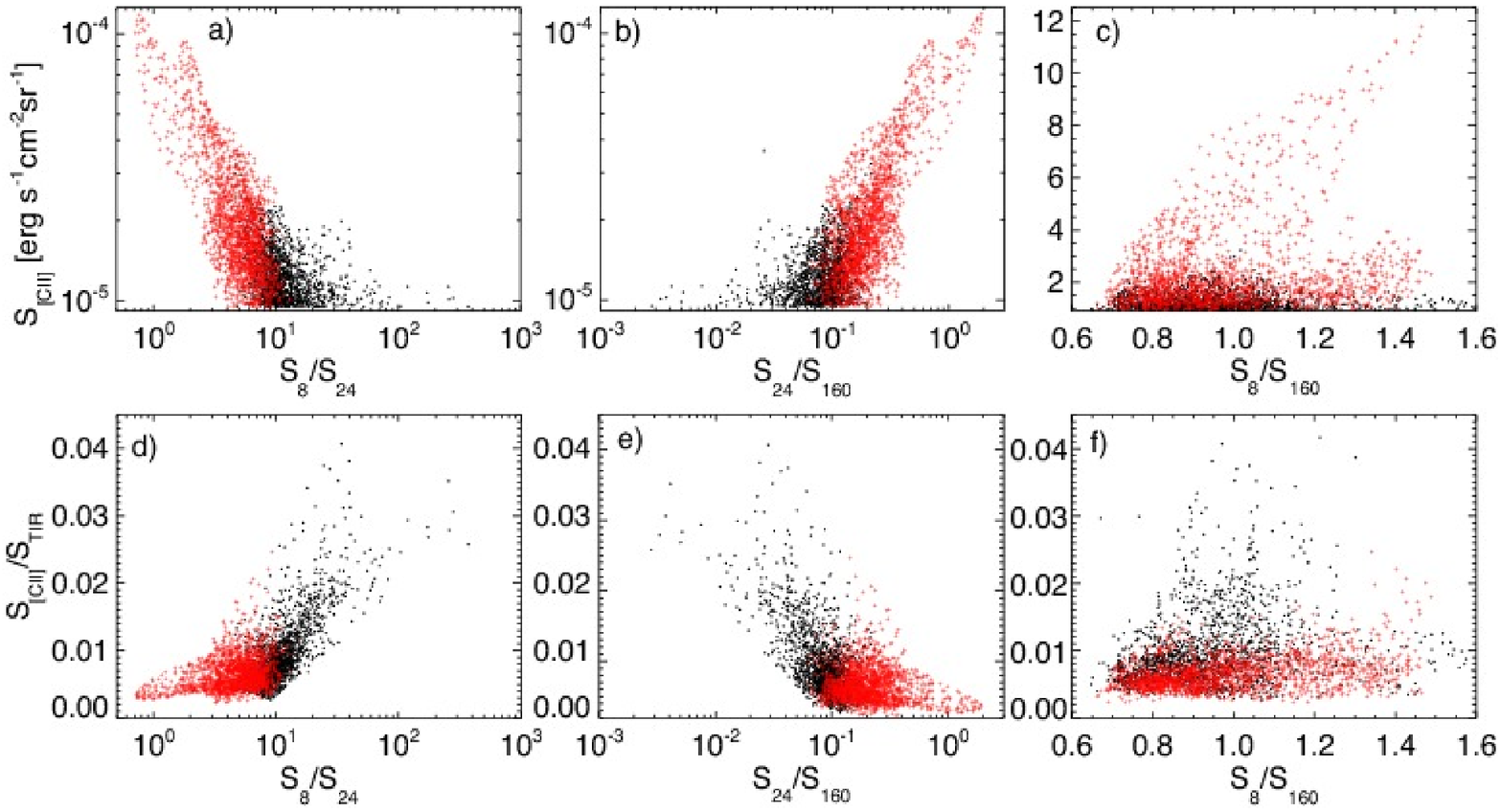} 
  \caption{[\cii] surface brightness (top) and the
    $\mathrm{S_\mathrm{[\cii]}/S_\mathrm{TIR}}$ (bottom) as a function
    of grain component emission ratios as traced by the Spitzer bands.
    The black dots and red crosses denote the different diffuse and SF
    pixels, respectively. Note, that the high values in the bottom
    three panels are due to noise in the [\cii] map (See text for
    details). See online version for colour.}
  \label{fig:combined_figure} 
\end{figure*}

The 8, 24, 70 and 160$\mu$m bands used in this study trace the
emission from distinct grain populations. We use these measurements to
study [\cii] cooling and PE heating with respect to grain abundance.
We adopt the generally accepted interpretation of the dust
constituents and at which wavelength they emit
\citep[e.g.][]{desert:1990,draine:2001}. The 8$\mu$m band is dominated
by PAH emission. The 24$\mu$m band mainly traces the emission of
stochastically heated carbonaceous grains with sizes less than
$\sim0.01\mu$m; termed very small grains (VSGs). The 160$\mu$m band
mainly traces the grains larger than $\sim0.01\mu$m, in radiative
equilibrium; termed big grains (BGs). Finally, the emission detected
in the 70$\mu$m band probably represents a combination of the BG
continua and the VSG emission, and may also trace grains
  stochastically heated, thus not in thermal equilibrium with the
  interstellar radiation field. Hereafter, we represent the PAH, VSG
and BG components with 8, 24 and 160 $\mu$m band, respectively.

The spatial variation of PAH emission in galaxies has been studied
extensively \citep[e.g.][]{roche:1985, aitken:1985, leach:1987,
  voit:1992, cesarsky:1996, verstraete:2001, siebenmorgen:2004,
  povich:2007}. The studies consistently find a lack of PAH emission
for the most active regions, i.e., \hii~regions, starburst galaxies
and AGNs. An interpretation is that the PAHs in these regions are
destroyed due to the hard, intense radiation field. The VSG grain
emission peaks in the \hii~region while the PAH emission peaks in the
adjacent PDR.

We present ratio maps of $\mathrm{S_{8}}$, $\mathrm{S_{24}}$,
$\mathrm{S_{70}}$ and $\mathrm{S_{160}}$ to $\mathrm{S_\mathrm{TIR}}$
in Fig.~\ref{fig:ratio_maps}a-d. The grey lines enclose the SF regions
as given by our H$\alpha$ criterion. Consistent with the studies
mentioned above, Figs.~\ref{fig:ratio_maps}a,b show that the relative
PAH emission is weak in the SF regions, while the VSG emission peaks
in the SF regions. Fig.~\ref{fig:ratio_maps}d shows that the BG
emission relative to TIR mostly peaks in the diffuse medium. BGs may
be present throughout the \object{LMC}, the high flux of UV photons in
the \hii~regions entails that they are hotter in these regions and
therefore radiate at wavelengths blueward of the 160$\mu$m band.
Finally, we note that the 70$\mu$m emission relative to TIR
(Fig.~\ref{fig:ratio_maps}c) most closely follows that of the VSG
emission.

\subsection{Grain Component Emission Relative to [\cii] Emission}
To study how the [\cii] emission varies with grain population, we
examine this parameter as a function of the ratios of grain component
emission. It should be noted, though, that the ratios represent ratios
of grain \textit{emission} and not that of grain abundances.

Fig.~\ref{fig:combined_figure} presents $\mathrm{S_\mathrm{[\cii]}}$
as a function of $\mathrm{S_{8}/S_{24}}$, $\mathrm{S_{24}/S_{160}}$,
and $\mathrm{S_{8}/S_{160}}$. We omit discussion of ratios with the
70~$\mu$m band as they exhibit the same behaviours as the ratios with
the 24~$\mu$m band. We note the following trends:

\begin{itemize}
\item In the SF regions where the [\cii] emission is highest, the PAH
  emission is low relative to the VSG emission
  (Fig.~\ref{fig:combined_figure}a), which is explained by the fact
  that the PAHs do not survive in \hii~regions where the VSGs
    are known to peak. As the $\mathrm{S_{8}/S_{24}}$ ratio
  increases, the pixels trace more and more of the diffuse medium
  where the [\cii] emission is the lowest.
\item Panel b shows again that those regions with very prominent
  24~$\mu$m emission, i.e. SF regions, show strong [\cii] emission.
\item Fig.~\ref{fig:combined_figure}c shows that there is no
  correlation between the [\cii] emission and the
  $\mathrm{S_{8}/S_{160}}$ ratio for the diffuse medium. Some fraction
  of the SF pixels, though, show a rough increase with [\cii] as a
  function of $\mathrm{S_{8}/S_{160}}$. The regions with the most
  intense radiation fields will heat the surrounding grains to very
  hot temperatures so that the BG emission will shift to bluer
  wavelengths, outside of the 160$\mu$m filter. Toward the SF regions,
  the amount of PAH emission also decreases (see above). The decrease
  in the $160\mu$m band is greater than the decrease in the PAH
  emission toward the most intense SF regions. This distinction
  between the brightest and the fainter SF regions is not reflected in
  the relative [\cii] strength (Fig.~\ref{fig:combined_figure}e).
\end{itemize}

\subsection{Theoretical description of PE}
\label{sec:PE_heating_and_eff}
PE heating occurs as follows: absorption of a far-ultraviolet (FUV)
photon by an interstellar dust grain liberates an electron within the
grain. The electron travels through the grain, escapes, and then
overcomes any Coulomb attraction if the grain is charged. If the
absorbed energy exceeds the work function of the grain plus its
Coulomb potential, the electron escapes with excess kinetic energy.
That energy goes into heating the ISM via collisions with the gas
species. Because of this Coulomb potential, PE heating efficiency is
highly dependent upon the charge state of the grain. This, in turn, is
dependent upon the physical conditions which determine grain
ionisation and recombination rates. The PE efficiency ($\epsilon$) is
defined as:
\begin{equation}
  \label{eqn:definition_of_epsilon}
  \mathrm{\epsilon\equiv\frac{\Gamma_{PE}}{P_{abs}}},
\end{equation}
where $\mathrm{P_{abs}}$, represents the power absorbed by the grains
and $\mathrm{\Gamma_{PE}}$ is the photoelectric heating rate. Thus,
$\epsilon$ is the fraction of the power absorbed by the grains that
goes into heating the ISM.

Photoelectric efficiencies as a function of environmental conditions,
grain size distributions and grain compositions were calculated
semi-empirically by \citet{dejong:1977}, and \textit{ab initio}
calculations have been made by \citet{weingartner:2001} and
\citet{bakes:1994}.

Assuming an MRN grain distribution and only carbonaceous grains,
\citet{bakes:1994} derive an analytic expression for PE efficiency
(valid for gas temperatures much less than 10\,000~K):
\begin{equation}
  \label{eqn:b_and_t_eff}
  \mathrm{\epsilon(G_{o},T,n_{e})}=\mathrm{\frac{4.87\cdot10^{-2}}{1+4\cdot10^{-3}\gamma^{0.73}}}~,
\end{equation} 
where $\gamma$ is the ratio of ionisation to recombination rate
($\mathrm{{G_{o}T^{1/2}}{n_{e}}^{-1}}$). Thus, high ionisation rates
will increase grain charge and therefore lower efficiency.

\subsection{A Proxy for $\epsilon$}
\label{sec:proxy_for_ep}
$\mathrm{L_\mathrm{[\cii]}/L_\mathrm{TIR}}$ is a proxy for $\epsilon$
given several assumptions: {\it i)} that PE heating and the
$\mathrm{C^{+}}$ transition dominate the heating and cooling processes
respectively, i.e. $\mathrm{\Gamma_{PE}=\Lambda \cong
  L_\mathrm{[\cii]}}$ (where $\Lambda$ represents the cooling rate)
and {\it ii)} And that interstellar grains and molecules re-radiate
all of the energy absorbed in the infrared. However, the [\cii] line
is not always the dominant coolant. Therefore, to observationally
estimate $\epsilon$, some authors include the measured luminosities of
other FIR lines \citep[e.g.][]{meixner:1992, youngowl:2002}. We do not
have maps of the \object{LMC} in other FIR lines, and thus can not
include their contribution in the cooling rate. Therefore, it should
be kept in mind that our calculations of $\epsilon$ represent a lower
limit on the actual values of PE efficiency, in particular in the
densest regions.

Here we estimate the contribution from other lines that may be missing
within the 15$^{\prime}$ beam. We perform this estimation in the
\object{30 Dor} region, as this is the region in the \object{LMC}
where the contributions of other FIR lines to the cooling rate should
be the most significant. We estimate the contribution from the
[\oi]~63$\mu$m line, the dominant cooling line in regions where [\cii]
cooling is suppressed. The \object{30 Dor} complex was measured in
[\cii] and [\oi]~(63$\mu$m) by \citet{poglitsch:1995} with a
55$^{\prime\prime}$ (FWHM) beam aboard the KAO. They found peak
intensities in [\cii] and [\oi] of $\mathrm{1\cdot10^{-3}
  erg~s^{-1}cm^{-2}sr^{-1}}$ and $\mathrm{6.1\cdot10^{-4}
  erg~s^{-1}cm^{-2}sr^{-1}}$ respectively. \citet{vermeij:2002b} also
measured these lines for several regions in \object{30 Dor} with the
LWS which has a beam 80$^{\prime\prime}$. They found that the [\oi]
intensity is about twice the [\cii] intensity.

To estimate the contribution of [\oi] in our 15$^{\prime}$ beam, we
assume two components within the beam: {\it 1)} the smaller region as
measured by \citet{poglitsch:1995}, and {\it 2)} more extended
emission. For 1, we use the values as measured by
\citet{poglitsch:1995}. For 2, we use the differences between the line
strengths as listed by \citet{vermeij:2002b} and listed by
\citet{poglitsch:1995}. We assume that the ratio of [\oi]/[\cii] in
the extended region holds throughout the 15$^{\prime}$ beam and scale
that to the full [\cii] measured in the large beam. This is a very
conservative estimate of the [\oi] contribution because it probably
overestimates the [\oi] line strength as the true ratio most likely
decreases with distance. We thus estimate the contribution of
[\oi]~63$\mu$m to be $\sim$20\% of the total gas cooling rate. This
number should be lower in other regions in the \object{LMC} since
\object{30 Dor} represents the most extreme SF region. We conclude
that, for our 15$^{\prime}$ beam, the [\cii] is representative of the
total gas cooling rate, but may slightly underestimate the cooling
rate in the most extreme cases.

Of the SF regions, \object{30 Dor} has the lowest efficiency as it is
the most intense SF region in the \object{LMC} (See
Tab.~\ref{tab:frac_of_CII}). In contrast the second brightest SF
region, \object{N11}, has a higher than average efficiency. The
regions defining \object{30 Dor} and \object{N11} were chosen by
constructing a rectangle centred on the brightest pixel in the
$24\mu$m image. The edges of the rectangles approximate three times
the mean S$_{24}$ level of the diffuse regions. We have included the
resulting rectangles in Fig.~\ref{fig:lmc_maps_1}a. One puzzling
result of this is that on average the diffuse regions exhibit a lower
efficiency than do the SF region. It should be noted that most of the
diffuse regions is faint and therefore affected more by the high noise
in the BICE map.

\subsection{The limited sensitivity of the BICE map: validity of observed efficiency variations}
\label{sec:simulations}
\begin{figure}
  \includegraphics[clip,width=8.8cm]{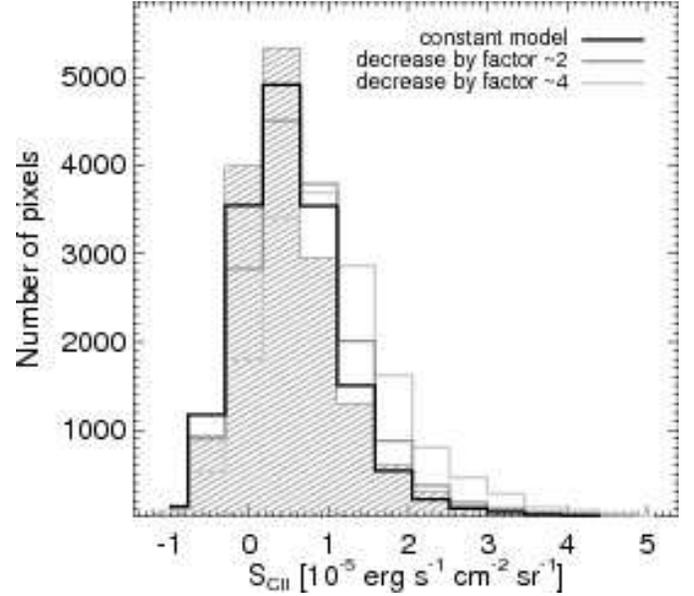}
  \caption{Fluxes as observed in the BICE maps versus those obtained
    from a simple simulation (section \ref{sec:simulations}). The
    figure shows the observed histogram of [\cii] surface-brightness
    (solid), the bin width is 1 $\sigma$ in [\cii]. The best fit model
    is shown in black (text for details). The dark grey line indicates
    the result of a simulation with a moderate decrease in efficiency
    as a function of TIR. The light grey line corresponds to a model
    with an decrease by a factor 4, the latter simulation is clearly
    inconsistent with the observations as it over predicts the number
    the number of bright pixels.}
  \label{fig:simulations} 
\end{figure} Before we can draw conclusions on the
  variations observed in the observed photo-electric efficiency
  parameter, we explored the possible existence of a systematic
variation of the efficiency simply as a function of G$_0$ by running a
set of simulations.The BICE map has limited sensitivity and as a
result a large part of the diffuse medium remains undetected or has
flux levels comparable to the noise level (See
Fig.~\ref{fig:map_efficiency}). Therefore, extra care must be taken in
deriving representative values of the [\cii] intensity in the diffuse
medium. In these simulations we assume that S$_\mathrm{TIR}$ is a
proxy for G$_0$. We simulate observed [\cii] maps by applying a
function of the form $\epsilon$=f(S$_\mathrm{TIR}$). The
S$_\mathrm{TIR}$ map is constructed using the Spitzer images and the
uncertainty in this map is negligible compared to the BICE map. After
this we add noise according to the noise level of the BICE map.
Fig.~\ref{fig:simulations} compares the observed histogram of [\cii]
values with several predicted histograms obtained following the
simulation outlined above. We have simulated maps using functions of
two forms {\it linear)} f(S$_\mathrm{TIR}$) = $a$ +
$b$*S$_\mathrm{TIR}$ and {\it power-law)} f(S$_\mathrm{TIR}$) =
$a$*S$_\mathrm{TIR}^p$. We determine the parameters which best
approximate the distribution of the observed S$_\mathrm{[\cii]}$
distribution. Some results of these simulations are shown in
Fig.~\ref{fig:simulations}. The closest match is obtained when
assuming a {\it constant} efficiency across the full range of G$_0$.
The data do not exclude modestly higher efficiencies in the faintest
regions by about a factor of two,. Steeper gradients are excluded,
since they clearly over-predict the number of pixels with values
between 3 and 5 $\sigma$ in the BICE map. At a given value of G$_0$ we
do find a significant spread of S$_{[\cii]}$/S$_\mathrm{TIR}$ in our
simulation. The spread, which is independent of S$_\mathrm{TIR}$, has
a magnitude of roughly a factor of two. The main conclusions of the
simulations are as follows:
\begin{itemize}
\item The value of S$_{[\cii]}$/S$_\mathrm{TIR}$ is independent of
  S$_\mathrm{TIR}$ over the range from 3\,10$^{-4}$ to 3\,10$^{-2}
  erg~s^{-1}cm^{-2}sr^{-1}$ .
\item The mean value of $\epsilon$ across the LMC is 0.45 \%
\item There is a spread around this value between 0.3 and 0.6
\item There is a  modestly higher value of S$_{[\cii]}$/S$_\mathrm{TIR}$, up to
  $\sim$ 1\% in the faintest regions 
\item The only region exhibiting a systematically lower value is the
  \object{30 Dor} region with a mean value of
  S$_{[\cii]}$/S$_\mathrm{TIR}$, of 0.35 \%.
\end{itemize}

\section{[\cii] Emission, Photoelectric Efficiency and Radiation Field
  in the LMC}
\label{sec:pe_eff_and_phys_cond}
\begin{figure}[!tp]
  \includegraphics[clip,width=8.8cm]{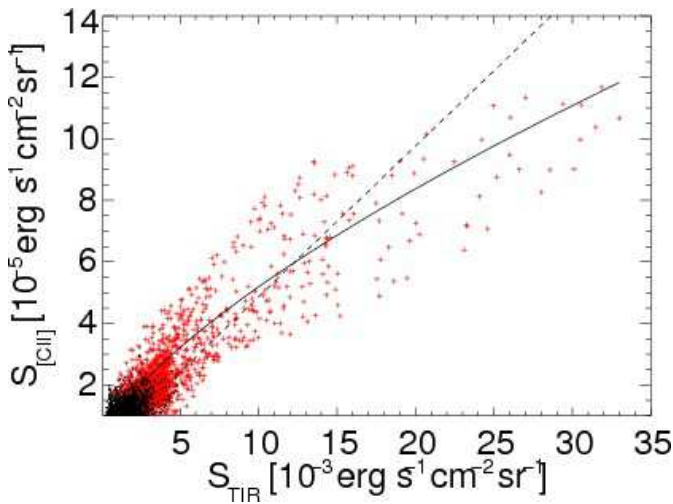}
  \caption{$\mathrm{S_\mathrm{[\cii]}}$ plotted as a function of
    $\mathrm{S_\mathrm{TIR}}$ for the diffuse and SF pixels. The solid line
    is a power law to the data. The dashed line is a linear fit to the
    data with the y intercept set to zero. Colours and symbols as in
    Fig.~\ref{fig:eff_vs_T_and_Halpha}. Colours and symbols as in
    Fig.~\ref{fig:eff_vs_T_and_Halpha}, see online version for colour.
  }
  \label{fig:L_Cii_vs_L_TIR} 
\end{figure}
\begin{figure*}[!tp]
  \includegraphics[clip,height=\textwidth,angle=90]{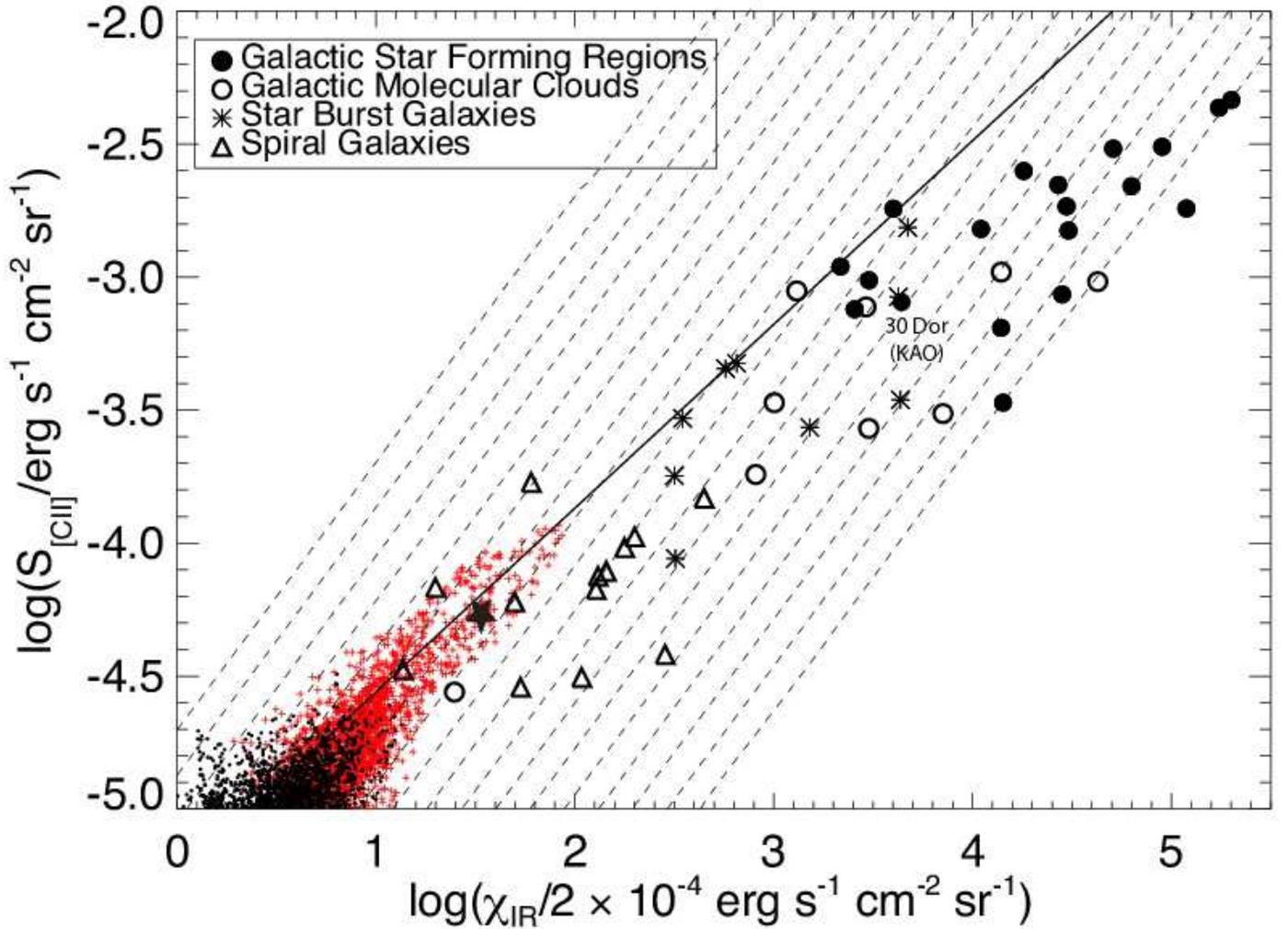}
  \caption{Diffuse and SF data plotted over Fig. 17 from
    \citet{stacey:1991}. .The solid line is a power law fit to our
    \object{LMC} data. Tracks of constant PE efficiency are also
    plotted (dashed lines). The first track on the left has an
    efficiency of 5\% and the last track on the right has an
    efficiency of $5\cdot 10^{-3}\%$. The interval between each
    line is a step of 1.5. Colours and symbols as in
    Fig.~\ref{fig:eff_vs_T_and_Halpha}, see online version for
    colour.}
  \label{fig:stacey_fig} 
\end{figure*}
Here, we examine the variations in the observed [\cii] emission and PE
efficiency as functions of radiation field.

\subsection {$\mathrm{S_\mathrm{[\cii]}}$ and Radiation Field}
\label{sec:cii_and_rad_field}
We can take $\mathrm{S_\mathrm{TIR}}$ as a proxy for the UV radiation
field in the case that most of the power absorbed by the dust is in
the UV, and the dust radiates isotropically. We plot
$\mathrm{S_\mathrm{[\cii]}}$ as a function of
$\mathrm{S_\mathrm{TIR}}$ in Fig.~\ref{fig:L_Cii_vs_L_TIR}. The [\cii]
emission increases as a function of $S_{TIR}$ and flattens at the
highest radiation fields, which is clearly seen in the power-law fit.
This flattening is dominated by the low efficiency observed in
\object{30 Dor}. We also tried to fit a straight line to
the data with the y intercept set at zero. The data is clearly better
described by the power law. The difference between the power-law and
the linear fit at the highest values of $\mathrm{S_\mathrm{TIR}}$ is
30\%, which is more than we can comfortably explain by
  missing line emission from other cooling fine structure lines
(Sec.~\ref{sec:proxy_for_ep}). The power-law fit yields the
  following prescription for the [\cii] surface brightness as a
  function of total infrared surface brightness throughout the LMC:
\begin{equation}
  \label{eqn:CII_as_function_of_TIR}
  \mathrm{S_\mathrm{[\cii]} = 1.25 \cdot 10^{-3}~{S_\mathrm{TIR}}^{0.69}},
\end{equation}
where surface brightness is given in
$\mathrm{erg~s^{-1}cm^{-2}sr^{-1}}$. The equation is valid for
$\mathrm{S_\mathrm{TIR}}$ between 3.2 and 33 $\cdot$
$\mathrm{10^{-4}erg~s^{-1}cm^{-2}sr^{-1}}$, the range from which the
fit was made. 

We interpret the flattening at the highest TIR in
Fig.~\ref{fig:L_Cii_vs_L_TIR} as a decrease in the PE heating rate. In
high radiation field and high temperature environments, grain charging
effects become important, the PE heating efficiency decreases and thus
the line cooling drops. This flattening is also observed in KAO data
of Galactic and extragalactic regions by \citet{stacey:1991}. In Fig.
17 of their paper, they plot $\mathrm{S_\mathrm{[\cii]}}$ as a
function of the FIR surface brightness (which they call
$\mathrm{\chi_{FIR}}$) normalised to $\mathrm{2 \cdot
  10^{-4}erg~s^{-1}cm^{-2}sr^{-1}}$. We show our data, their data and
Eq.~\ref{eqn:CII_as_function_of_TIR} in Fig.~\ref{fig:stacey_fig}. We
have converted our TIR values into $\mathrm{\chi_{FIR}}$ by assuming a
factor of two for TIR to FIR as given by \citet{hunter:2001}. We also
plot tracks of constant PE efficiency in Fig.~\ref{fig:stacey_fig}
(dashed lines) from $\epsilon =5\%$ (left) to $5 \cdot 10^{-3}\%$
(right); where $\epsilon = 5\%$ is chosen because it is close to the
highest efficiency in the theory of \citet{bakes:1994}.

Our data throughout the LMC follow the trend of
  \citet{stacey:1991} and extends to the lower left portion of the
plot. To understand why our SF points do not occupy the upper right
portion of the graph, one must consider that our beam size
($\mathrm{\sim 200pc}$) undoubtedly entails considerable mixing of the
phases of the ISM. \citet{stacey:1991} include measurements of
\object{30 Dor} (indicated on Fig.~\ref{fig:stacey_fig}). We use this
to gauge how beam size affects this figure. The measurement of
\object{30 Dor} from our data is indicated with a star. \object{30
  Dor} in the BICE beam has significantly lower IR and [\cii] surface
brightnesses, by factors of about 125 and 15 respectively. Note, that
because of the much smaller beam of the KAO, and because the Galactic
regions are closer, the data of \citet{stacey:1991} probe much smaller
spatial scales. On these scales, the contribution to cooling from the
[\oi] line as compared to the [\cii] line might be important. If the
[\oi] line were included in the \citet{stacey:1991} data, the points
at the upper-right side of the diagram would move up. Even so,
inclusion of these other lines would most likely not be sufficient to
move the data points up to the efficiency of the LMC, as this would
require them to be moved up the y-axis by at least an order of
magnitude. Such cooling contributions from other lines, even in the
most intense regions, are not expected. As with the decreasing trend
of efficiency in the Stacey data, we note that for our data,
\object{30 Dor} shows the most prominent decrease in efficiency.
Perhaps this reflects the fact that \object{30 Dor} is so bright that
it dominates the emission even in the large beam.

One must note though, that the other variables which control the PE
heating (i.e. gas temperature and electron density) are not constant
throughout Figs.~\ref{fig:L_Cii_vs_L_TIR} and~\ref{fig:stacey_fig}.
Perhaps the considerable spread in [\cii] at any given
$\mathrm{S_\mathrm{TIR}}$ in Fig.~\ref{fig:L_Cii_vs_L_TIR} and can be
attributed to those other variables.

\section {Correlation Between Radiation Field and Electron Density}
\label{sec:model_for_eff}
\begin{figure}[!tp]
  \includegraphics[clip,width=8.8cm]{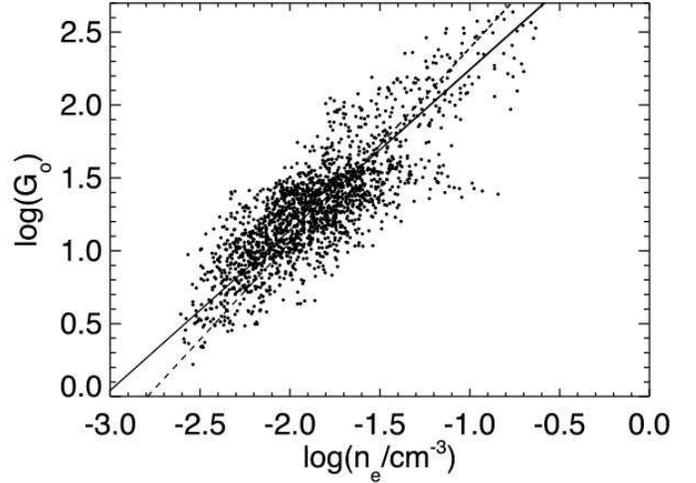}
  \caption{$\mathrm{G_{o}}$ as a function of $\mathrm{n_{e}}$ for the
    SF pixels as estimated from Eqns.~\ref{eqn:b_and_t_eff}
    and~\ref{eqn:S_TIR_and_radiation_field}. The solid and dashed
    lines are fits of the equation $G_{o} = \xi n_{e}^{p}$. The solid
    line is fit with both $\xi$ and p as free parameters and results
    in $\xi = 2,200$ and p = 1.1. The dashed line is a fit with p set
    to 4/3 and $\xi$ as a free variable. The result is $\xi =5,300$.}
  \label{fig:go_vs_ne}   
\end{figure}
\begin{figure}[!tp]
  \includegraphics[clip,height=8.8cm,angle=90]{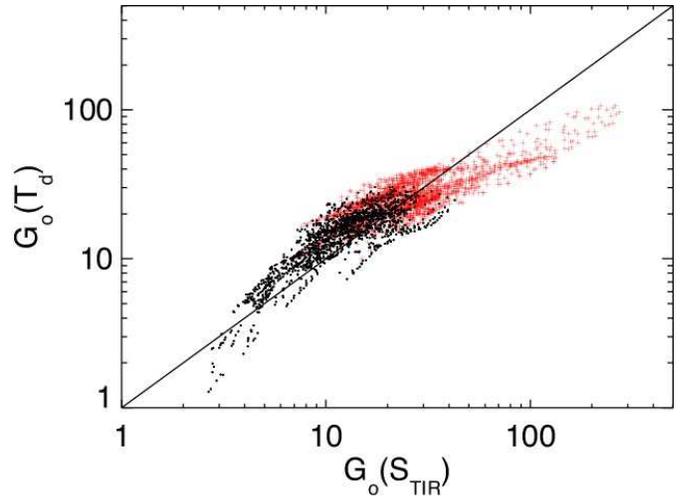}
  \caption{Values of $G_o$ for each pixel derived from the dust
    temperature assuming graphitic grains for the dust composition
    versus values of $G_o$ given by
    Eq.~\ref{eqn:S_TIR_and_radiation_field}. The solid line denotes
    agreement between the two methods. Colours and symbols as in
    Fig.~\ref{fig:eff_vs_T_and_Halpha}, see online version for
    colour.}
  \label{fig:G_o_vs_G_o} 
\end{figure}
\begin{figure}[!tp]
  \includegraphics[clip,width=8.8cm]{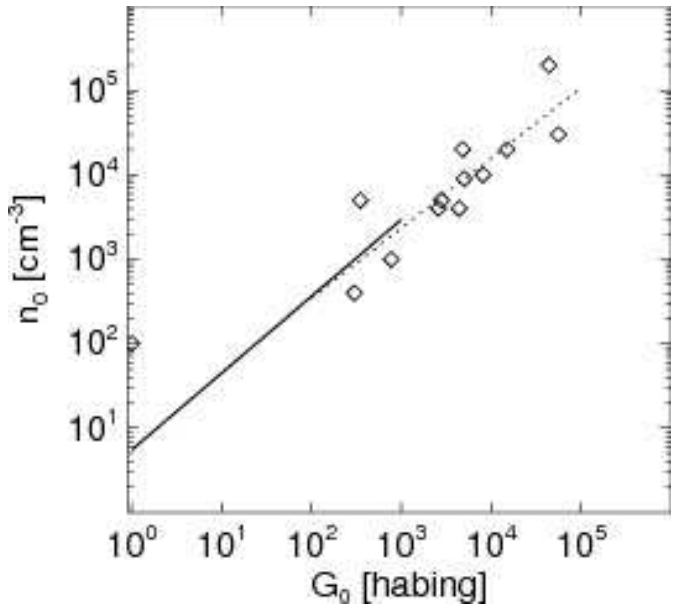}
  \caption{Comparison of the interrelation between $G_o$ and $n_0$
    that we find and that is found within galactic PDR regions by
    \citet{youngowl:2002}. The diamonds indicate the data from
    \citet{youngowl:2002}, while the dashed line is their best
    fit. The drawn line shows $n_0$ as a function of $G_0$ for each
    fixel in our analysis. Here $n_0$ is derived from $n_e$
    (Fig.~\ref{fig:go_vs_ne}) assuming that the electron density
    equals the carbon density.}
  \label{fig:compare_young_owl} 
\end{figure}
In this section we use the data and the physics of PE heating to
derive values for $\mathrm{G_{0}}$ and $\mathrm{n_e}$. We find that
the observed constant efficiency translates into a tight correlation
between the density of the radiation field and the density of
electrons. We discuss the physical interpretation of this correlation
and in particular we compare our results with those of
\citet{youngowl:2002} and \citet{malhotra:2001}.

To first order the observed efficiency is constant throughout the LMC.
This constancy of 0.45\% translates into a constant value of $\gamma
\approx 4\,10^{4}$ (Eq.~(\ref{eqn:b_and_t_eff}). A typical value for
PDR gas temperature is $\mathrm{T\sim 300K}$. To account for the
contribution from the diffuse ISM \citep[$\mathrm{T\sim
  50-100K}$][]{tielens:2005}) we adopt a value of T = 75\,K. Assuming
an average temperature of 75~\,K across the SF pixels we find that all
of the LMC regions, averaged over our 225 pc regions here, are
typified by G$_0$/n$_e$ $\approx$ 5000.

Several explanations for the constancy of the observed efficiency and,
as a consequence, the derived interrelation between G$_0$ and n$_e$
come to mind.

{\it i)} If, in our large beam, the radiation field is dominated by
similar PDR regions, then the number of PDRs in that beam will
determine variations in the total radiation field, but will not change
the intrinsic PE efficiency.

{\it ii)} If, in our large beam, the light is dominated by emission
from the diffuse medium, PE efficiency should remain fairly constant
as the physical conditions within this medium are relatively
invariant.

{\it iii)} The value of G$_o$/n$_e$ is constant throughout the
different types of media that make up the LMC. In fact, for several
Galactic PDR regions, \citet{youngowl:2002} do indeed find that G$_o$
and n$_e$ scale with each other. They studied the PDRs of a sample of
reflection nebulae. Their observations of FIR atomic fine structure
lines and the FIR continuum allowed them to obtain estimates of
radiation fields and gas densities (n$_0$).

It is interesting to compare our results in more detail with those
obtained by \citet{youngowl:2002} and \citet{malhotra:2001}. The
latter measured line emission ratios across entire galaxies for a
large sample of galaxies with varying morphologies. They used PDR
models to obtain values for radiation fields and gas densities. Both
then fitted power laws to their data. For a function of the form,
$\mathrm{G_{o} = \xi^{\prime}n_{o}^{p}}$, \citet{youngowl:2002} find
$\xi^{\prime} = 0.09$ and $\mathrm{p} = 1.2$, and
\citet{malhotra:2001} $\xi^{\prime} = 0.23$ and $\mathrm{p = 1.33}$
(estimated from Fig.~11 of \citet{malhotra:2001}). In other words,
they also find that the density scales roughly linearly with the
radiation density. 

Here, we estimate $G_0$ and $n_e$ for independent pixels in the LC. We
do this only for the SF-regions. One, because these are the brightest,
reliably detected regions and two, because the conversion from
S$_\mathrm{TIR}$ to G$_0$ uses the assumption of central illumination
which is more likely to hold in those regions. We use the observed
efficiency to invert the efficiency equation of
\citet[][Eq.~(\ref{eqn:b_and_t_eff})]{bakes:1994}. This yields values
of $\gamma$ for each pixel. We measure the G$_0$ using
S$_\mathrm{TIR}$, and assume a reasonable value for $T$. Thus we can
derive values of n$_e$ for each pixel.

Following the examples of \citet{meixner:1992, youngowl:2002,
  steinman:1997} we use the measured TIR surfaces brightnesses and
assume a certain geometry in order to estimate values of
$\mathrm{G_{o}}$. Assuming {\it 1)} that the illuminating sources are
at the centre of each pixel, {\it 2)} that the dust resides at edges
of the pixels, {\it 3)} 100\% conversion of FUV radiation to IR
radiation, {\it 4) }and that the dust radiates isotropically, then
\begin{equation}
  \label{eqn:S_TIR_and_radiation_field}
  G_{o}=C_{\mathrm{S\rightarrow G}}\,S_{\mathrm{TIR}},
\end{equation}
with $C_{\mathrm{S\rightarrow G}}= 4 \,10^{-4}$, $G_{o}$ in units of
the one dimensional habing field (1.6\,10$^{-3}$
ergs\,s$^{-1}$\,cm$^{-2}$) and $S_{\mathrm{TIR}}$ in
ergs\,s$^{-1}$\,cm$^{-2}$. Other geometries; like the dust residing
halfway between the source and the edge, homogeneous distribution of
sources and dust or uniformly illuminated dust; changes
$\mathrm{C_{S\rightarrow G}}$ only by a small amount. The $G_0$ thus
calculated is shown in Fig.~\ref{fig:go_vs_ne} versus the derived
electron density. We find a best fit power-law of the form $G_0
[habing] = 2\,200 {n_e}^{1.1} [cm^{-3}]$. Note, that the results of
\citet{malhotra:2001} and \citet{youngowl:2002} are expressed in
n$_{o}$ while we use n$_e$ and thus they differ by the ionisation
fraction.

In order to numerically compare our result we convert $n_e$ to $n_0$
assuming that the average electron density is dominated by
diffuse-ish, outside of the H~{\sc ii}~regions. regions. In this case
the prime donor of free electrons is carbon which is the most abundant
element with an ionisation potential (11.3 eV) below that of hydrogen
(13.6 eV). Taking into account the metallicity of the LMC we find a
conversion factor (H/C) from $n_e$ to $n_0$ of 6000. It can be seen in
Fig.~\ref{fig:compare_young_owl} that our values follow the trend
established by \citet{youngowl:2002} well.

\citet{youngowl:2002} and \citet{malhotra:2001} propose that the
correlation can be physically interpreted as a balance of gas
pressures between the PDR and the \hii~region. \citet{youngowl:2002}
present a simple analytical model assuming this balance of pressures
and they find that $G_{o}$ should scale with $n_{o}^{4/3}$. For both
\citet{youngowl:2002} and \citet{malhotra:2001}, the power of 4/3 fits
to within the uncertainties of their data and fits within the envelope
of the uncertainty of our data (Fig.~\ref{fig:go_vs_ne}). We have
performed an F-test to determine whether the fit with $p$ as a free
parameter is an improvement compared to strict pressure balance (See
Fig.~\ref{fig:go_vs_ne}). The shallower power-law ($p$=1.1) does
decrease the reduced $\chi^2$ significantly although it remains well
above unity.

The cause behind the correlation is unclear, especially one that would
hold over such a range of scales. It is difficult to envisage
pressure-balance between the powering \hii~regions and the 200 ~pc
regions of space that we are sampling. This is even more true for the
results of \citet{malhotra:2001}, who sample entire galaxies and still
find a strong correlation between the average radiation field and the
average density. Such a pressure balance would work if each single
region or galaxy is dominated by a single (or at most e few)
cluster(s) of young stars that cause a single prominent PDR to plough
into the containing molecular cloud.

\subsection {Dust Temperature in the LMC}
\label{sec:determine_dust_temp}
In the previous sections, we used the measured TIR surface brightness
as a proxy for radiation field. There could be a worry, though, that
$\mathrm{S_\mathrm{TIR}}$ can vary without actual variations
  in the  radiation field. This may be due to differences in the
amount of and characteristics |emitting material along the
line of sight as a result of varying densities or a varying thickness
of the \object{LMC}. We explore these concerns by comparing
$\mathrm{G_o}$ as derived from $\mathrm{S_\mathrm{TIR}}$ to
$\mathrm{G_o}$ as derived by a rough calculation of the dust
temperature as measured by the 70~$\mu$m to 160~$\mu$m ratio. This
ratio in not affected by varying amounts of material along the
line-of-sight.

We derive a dust temperature indicator assuming modified black-body
radiation with an emissivity of the following functional form: $Q
\propto \nu^{\beta}$, with $\beta=1.5$. This is a crude approximation,
since in reality dust grains do not emit as modified black-bodies. The
dust temperature will vary with grain-size and also with grain
composition. Moreover, a part of the emission we observe at 70~$\mu$m
is not due to dust at an equilibrium temperature (See
Sec.~\ref{sec:dist_of_grain_emission}), but arises from
stochastically heated grains. For these reasons, the absolute value of
the derived temperature is not very reliable, but the relative
temperatures are. In spite of the above caveats, the temperatures that
we find are all-together reasonable and compare well with the results
of more in-depth studies. The mean dust temperature we find
(${\mathrm{\overline{T_{d}} = 24.0 \pm 2.2K}}$) is consistent with the
ones found by \citet{sakon:2006} ($23.9\,K$) and \citet{aguirre:2003}
($25.0 \pm 1.8\,K$).

Using the relationship between $\mathrm{G_o}$ and $\mathrm{T_d}$ given
by \citet{tielens:2005} for graphite grains, we calculate
$\mathrm{G_o(T_d)}$ for each pixel. We plot
$\mathrm{G_o(S_\mathrm{TIR})}$ versus $\mathrm{G_o({T_d})}$ for each
pixel in Fig.~\ref{fig:G_o_vs_G_o}. The solid line denotes perfect
correspondence. The correspondence between the two G$_0$ indicators is
clear and thus we conclude that S$_\mathrm{TIR}$ is indeed a valid
proxy for G$_0$.

\section{Photoelectric Efficiency and Grain Distribution in the LMC}
\label{sec:dust_and_PE_heating}
\begin{table}[!tp]
  \caption{Results of Efficiency Calculations}
  \begin{tabular}{l||ccccc}
    \hline
    \hline
    \multicolumn{1}{c}{Region}  &$\epsilon_{8}$& $\epsilon_{24}$& $\epsilon_{70}$&$\epsilon_{160}$& $\chi^{2}$\\
    \hline
    Integrated over \object{LMC} & 0.02       & 0.00           & 0.02          & 0.01         & 1.1\\
    SF regions                   & 0.06       & 0.00           & 0.00          & 0.01         & 2.0\\
    \object{30 Dor}              & 0.05       & 0.00           & 0.00          & 0.01         & 4.7\\
    \object{N11}                 & 0.05       & 0.04           & 0.00          & 0.00         & 2.8\\
    \hline
    \it Diffuse regions              & \it 0.00       & \it 0.11           & \it 0.02          & \it 0.02         & \it 0.8\\
    \hline
    \hline
  \end{tabular}
  \label{tab:eff_calc_1}
\end{table}  
Theoretical studies have shown that the extent of grain charging and
therefore the efficiency of PE heating is not only dependent on
environmental conditions, but also on grain size and grain species.
Interstellar molecules such as PAHs and the smallest grains contribute most extensively to the PE heating \citep{watson:1972,
  bakes:1994, weingartner:2001}. For example, \citet{bakes:1994} have
found that approximately half of the heating is from grains with less
than about 1\,500 C-atoms ($\sim$ 15\,\AA).

Qualitatively, the scaling of efficiency with grain size is due to the
fact that the ionisation rate is approximately $\mathrm{\propto
  N_{C}}$ since the FUV absorption cross section is approximately
$\mathrm{\propto N_{C}}$. But, the recombination rate increases more
slowly with the number of carbon atoms as it scales with the grain
size (about $\mathrm{N_{C}^{1/3}}$) . Therefore, the fraction of
ionised grains increases with grain size, scaling by about
$\mathrm{N_{C}^{2/3}}$.

There has been a plethora of theoretical studies on grain species and
PE heating, but relatively few observationally based studies.
Therefore, we now examine PE efficiency as a function of grain species
by using each band as a tracer of grain abundance.

\subsubsection{A Calculation of Efficiencies}
\label{sec:grain efficiencies}
Authors usually take [\cii]/FIR as a general efficiency, encompassing
the contribution to PE heating of all constituent grain populations.
But, as each grain component is expected to have different intrinsic
PE efficiencies, it is interesting to isolate the contribution of PE
heating from the various species. Therefore, to quantify the
importance of the various grain species to the PE heating process, we
follow the example of \citet{habart:2001}.

Using IRAS data, \citet{habart:2001} quantified the amount of emission
attributed to the PAH, VSG and BG populations in the $\rho$ Opiuchi
complex. Using ISO observations of the [\cii], [\oi] and
$\mathrm{H_2}$ lines, they then calculated $\mathrm{\epsilon_{PAH}}$,
$\mathrm{\epsilon_{VSG}}$, $\mathrm{\epsilon_{BG}}$ with a linear
combination fit of their gas cooling rate to their grain emission
rates (further detail given below). They found that the PAH population
is attributed with the highest photoelectric efficiency, while the BG
population is attributed with the lowest. They found
($\mathrm{\epsilon_{PAH}}= 3\%$, $\mathrm{\epsilon_{VSG}}=1\%$,
$\mathrm{\epsilon_{BG}}=0.1\%$).

To perform this calculation, we start with the definition of
efficiency for a certain grain population, j, which is given as
\begin{equation}
  \label{eqn:habart_1}
  \mathrm{\epsilon_{pop~j}\equiv \frac{\Gamma_{PE}(pop~j)}{P_{abs}(pop~j)} \cong \frac{S_\mathrm{[\cii]}(pop~j)}{S_{emitted}(pop~j)}}.
\end{equation}
The substitutions of $\mathrm{\Gamma_{PE}}$ and $\mathrm{P_{abs}}$
with $\mathrm{S_\mathrm{[\cii]}}$ and $\mathrm{S_{emitted}}$ were
already discussed in Sec.~\ref{sec:proxy_for_ep}. To solve for the
total amount of emitted [\cii], we multiply both sides of
Eq.~\ref{eqn:habart_1} with the denominator on the right side of the
equation, and then sum over all grain populations:
\begin{equation}
  \mathrm{S_\mathrm{[\cii]}=\sum_{all~pop}{\epsilon_{pop~j}S_{emitted}(pop~j)}}.
\end{equation}
Finally, if we consider the emission in the 8, 24, 70 and 160$\mu$m
bands as arising from distinct grain populations, and that these are
the main populations that contribute to PE heating, we obtain:
\begin{equation}
  \label{eqn:habart_3}
  \mathrm{S_\mathrm{[\cii]}\cong \epsilon_{8}S_{8}+\epsilon_{24}S_{24}+\epsilon_{70}S_{70}+\epsilon_{160}S_{160}}.
\end{equation}
The variable, $\epsilon_{8}$, should reflect the of PAH efficiency,
$\epsilon_{24}$ of the VSG efficiency, $\epsilon_{70}$ of the a
combination of the VSG and BG efficiency and $\epsilon_{160}$ of just
the BG efficiency.

To solve for the efficiencies of Eq.~\ref{eqn:habart_3}, we fit a
linear combination of the IRAC and MIPS data to the [\cii] map with a
$\chi^2$ minimisation. Further, we force the efficiencies to be
between 0 and 1. We perform this calculation across the whole galaxy,
for the \object{30 Dor} and \object{N11} regions and for the diffuse
and SF regions in the \object{LMC}. All calculated efficiencies, along
with the $\chi^2$ value for each fit are given in
Table~\ref{tab:eff_calc_1}.

Table~\ref{tab:eff_calc_1} shows that with the exception of
\object{N11}, $\epsilon_{24}$ and $\epsilon_{70}$ are always zero and
$\epsilon_{8}$ and $\epsilon_{160}$ are non-zero. We show the values
derived for the diffuse medium as well. However, these values should
be taken with extreme care, because of large contribution of noise to
these pixels. We have experimented with deriving grain-efficiencies
also using simulated [\cii] maps and find that the 2\% value for the
BG is found persistently but that the $\epsilon$-VSG grain is very
sensitive to the exact noise characteristics of the [\cii] map and can
not be trusted.

The values in Table~\ref{tab:eff_calc_1} show that $\epsilon_{8}$ is
greater than $\epsilon_{160}$ for every region considered. This is
quantitative proof that the PAH emission is more spatially correlated
with the [\cii] emission than the BG emission. This supports the
interpretation put forth in the previous section, that the PAH
population dominates the PE efficiency and plays an
  important role in the Photoelectric heating of the gas.

The results for \object{N11} obviously differ from the other regions
considered as $\epsilon_{160}=0$ and as and $\epsilon_{24}\neq0$. This
might reflect offsets between the [\cii] emission and the Spitzer
bands due to the distinctive asymmetry towards \object{N11} already
discussed in Sec.~\ref{sec:convolved_data}.

\section{Conclusion}
Using the MIR to FIR SAGE maps, and the BICE [\cii] map of the
\object{LMC}, we have, for the first time, conducted an observational
study of PE heating and [\cii] cooling in relation to spatially
resolved grain emission throughout the \object{LMC}.

Integrated throughout the entire \object{LMC}, the [\cii] line
accounts for $0.64 \pm 0.01\%$ of the total infrared ($\sim 1.2\%$ of
the FIR). Applying a correction for the pixels below
$2\sigma_\mathrm{[\cii]}$, we find that the [\cii] line accounts for
$1.32 \pm 0.01\%$ of the FIR. This value is greater than that of
normal and gas rich galaxies (with values typically from 0.1-1\%), as
found in other low metallicity galaxies.

Distinguishing environments by H$\alpha$ surface brightnesses and by
location, we find that the [\cii] line contributes significantly less
to the TIR emission in SF regions versus diffuse ISM regions:
\begin{itemize}
\item $\mathrm{L_\mathrm{[\cii]}/L_\mathrm{TIR}=0.57 \pm 0.01 \%}$ for the SF regions
\item $\mathrm{L_\mathrm{[\cii]}/L_\mathrm{TIR} = 0.42 \pm 0.01\%}$ for \object{30 Doradus}
\item $\mathrm{L_\mathrm{[\cii]}/L_\mathrm{TIR} =0.56 \pm 0.03\%}$ for \object{N11}
\item $\mathrm{L_\mathrm{[\cii]}/L_\mathrm{TIR}= 0.79 \pm 0.01 \%}$ for the diffuse regions. 
\end{itemize}
We also calculate the contribution of the total output of [\cii]
emission from the \object{LMC} from the same regions. We find that,
although the SF regions have the highest surface brightness values,
most of the \object{LMC}'s [\cii] emission originates from the diffuse
medium:
\begin{itemize}
\item $\mathrm{L_\mathrm{[\cii]}/L_{[\cii], LMC}= 31.5 \pm 0.1 \%}$ for the SF regions
\item $\mathrm{L_\mathrm{[\cii]}/L_{[\cii], LMC} = 5.58 \pm 0.1 \%}$ for \object{30 Doradus}
\item $\mathrm{L_\mathrm{[\cii]}/L_{[\cii], LMC} =1.53 \pm 0.1 \%}$for \object{N11}
\item $\mathrm{L_\mathrm{[\cii]}/L_{[\cii], LMC} = 68.5 \pm 0.1 \%}$ for the diffuse regions. 
\end{itemize}

To examine variations in PE efficiency within the \object{LMC}, we use
$\mathrm{S_\mathrm{[\cii]}}$ as a proxy for the total PE heating rate.
We estimate that this assumption is valid for our 15$^{\prime}$ beam
for all but the most active SF regions which might underestimate the
total heating rate by at most 20\%. We study how PE efficiency varies
with environment using the observed values of
$\mathrm{S_\mathrm{TIR}}$ as an indicator of the local radiation
field. As a function of $\mathrm{S_\mathrm{TIR}}$,
$\mathrm{S_\mathrm{[\cii]}}$ flattens at the highest value of
$\mathrm{S_\mathrm{TIR}}$. The flattening trend is interpreted as a
decrease in the PE efficiency for the most illuminated regions in the
\object{LMC}. We provide a prescription for
$\mathrm{S_\mathrm{[\cii]}}$ as a function of
$\mathrm{S_\mathrm{TIR}}$ (Eq.~\ref{eqn:CII_as_function_of_TIR}) by
fitting a power law to the data. Such a decrease in efficiency is
theoretically expected due to grain charging effects.

Previous studies have found a correlation between these two
parameters, with $\mathrm{G_{o} \propto n_{e}^{4/3}}$. This relation
has been explained by invoking a simple model assuming
pressure-balance between \hii~regions and the adjacent PDRs.
Theoretically, the PE efficiency depends strongly on the
recombination-rate, and thus on the ratio of $\mathrm{G_o/n_e}$. We
thus calculate values for $\mathrm{G_o}$ and $\mathrm{n_e}$ using the
observed efficiencies. We convert the observed
$\mathrm{S_\mathrm{TIR}}$ to $\mathrm{G_o}$, assuming illumination by a
central source in each 45$\times$45~pc pixel. We find that a similar
scaling-relation between $\mathrm{G_o}$ and $\mathrm{n_{e}}$ holds for
the \object{LMC}.It is unclear why the Str\"{o}mgren
sphere argument should hold on the large $\sim$45~pc scale that we
probe.

We analyse observed PE efficiencies in relation to the grain component
emission from each Spitzer band. We note that this is the first such
analysis utilising spatially resolved grain emission components
throughout an entire galaxy. From the correlations between observed
efficiency and the grain component emission, and from a calculation of
the PE efficiencies for each population, we show that the PAH emission
is the most spatially correlated with the PE heating rate. We
therefore conclude that it is the PAH population that dominates the PE
heating process.

The efficiency of the PE heating process is dependent on both
environmental conditions, such as radiation field, and grain
abundances. Our study has examined PE efficiency within the
\object{LMC} without fully disentangling the extent of PE heating due
to existence of grain populations favourable to PE heating and the
extent due to environmental conditions favourable to PE heating.
Disentangling the effects of both on the observed PE efficiency,
however, is difficult as the regions where PAH populations are
destroyed naturally have intense radiation fields which also suppress
the extent of PE heating. To break this degeneracy we will undertake
detailed SED modelling of the different regions in the LMC in an
upcoming paper. Using the SED models we can independently solve for
radiation field values and PAH abundances.

\begin{acknowledgements}
  We would like to thank F. Boulanger, L. Verstraete and A. Jones for
  their helpful conversations. Meixner, Vijh, Sewilo and Leitherer
  have been funded by the NASA/Spitzer grant 1275598, and NASA
  NAG-12595.
\end{acknowledgements}

\bibliographystyle{aa}
\bibliography{0968}
\end{document}